\documentclass[aps,floats,nofootinbib,amssymb,preprint,superscriptaddress]{revtex4-1}
\usepackage{textcomp}
\usepackage{amsmath}
\usepackage{amssymb}
\usepackage{tabularx}
\usepackage{graphicx}
\usepackage{inputenc}
\usepackage{url}
\usepackage{hyperref}
\usepackage{booktabs} 
\usepackage{array} 

\begin{document}

\title{LHC and dark matter signals of $Z'$ bosons}
\author{Vernon Barger}
\affiliation{Department of Physics, University of Wisconsin, Madison, WI 53706, USA}
\author{Danny Marfatia}
\affiliation{\mbox{Department of Physics \& Astronomy, University of Kansas, Lawrence, KS 66045, USA}}
\affiliation{Department of Physics, University of Wisconsin, Madison, WI 53706, USA}
\author{Andrea Peterson}
\affiliation{Department of Physics, University of Wisconsin, Madison, WI 53706, USA}


\begin{abstract}
We customize the simulation code FEWZ (Fully Exclusive W, Z Production) to study $Z'$ production at the LHC for 
both $\sqrt{s} = 8$~TeV and 14~TeV.  Using the results of our simulation for several standard benchmark $Z'$
models, we derive a semi-empirical expression for the differential cross section, that permits the determination of $Z'$ couplings in a model-independent manner. We evaluate cross sections and other observables for large classes of models, including the common $E_6$, left-right and $B-L$ models, as a function of model parameters. We also consider a hidden sector $Z'$ that couples to standard model fermions via kinetic and mass mixing and serves as a mediator of isospin-violating interactions with dark matter. We combine the results of LHC $Z'$ searches and dark matter direct detection experiments with global electroweak data to obtain mass-dependent constraints on the model parameters.
\end{abstract}

\maketitle

\section{Introduction}

A simple extension of the Standard Model (SM) is the addition of an extra $U(1)$ gauge symmetry, with associated neutral $Z'$ gauge boson. Extra $U(1)$ symmetries are a necessary part of many interesting new physics scenarios, including several Grand Unified Theories and string-inspired model constructions. Generic $Z'$ models can have many new physics features, including generation-dependent couplings, $Z-Z'$ mixing, and new fermions; see {\it e.g.}, Refs.~\cite{Langacker:2008yv, Carena:2004xs, Leike:1998wr, Hewett:1988xc, 
London:1986dk}. We first study the simplest models with generation-independent couplings and no $Z-Z'$ mixing. There are a number of such models that are theoretically relevant, such as the $E_6$ GUT models and the $B-L$ model.  In Section \ref{DMZ'}, we consider a model that includes mass and kinetic mixing between the $Z'$ and the $Z$~\cite{Frandsen:2011cg}, which has applications to isospin-violating dark matter 
scattering~\cite{Feng:2011vu}.

We expand the simulation code FEWZ 2.1~\cite{Gavin:2012kw, Gavin:2010az} to study the production and decay of $Z'$ bosons at the LHC through the process $pp\rightarrow Z' \rightarrow l^+l^-$. FEWZ includes up to NNLO in perturbative QCD and is fully differential in the lepton phase space. This allows for the precise calculation of the  $Z'$ cross sections and differential distributions with realistic experimental acceptances. In Sections \ref{detection} and \ref{Sim}, we briefly review $Z'$ detection and introduce several common benchmark models that we use to demonstrate the efficacy and validity of our simulation.

In Section~\ref{fits} we use FEWZ to derive a semi-empirical expression for the differential cross section $\frac{d\sigma}{dy \, d\cos{\theta}}$. We show that, with sufficient data, this formula can be used to determine the couplings of the $Z'$ and set limits on model parameters with good precision.

In Section~\ref{E6 analysis}, we apply our fit to the $E_6$ model class and the production of two $Z'$ bosons. The heavier of the two mass eigenstates is often assumed to be too heavy for collider detection, but we show that it could be accessible at the LHC for a certain range of mixing angles. Throughout, we focus on $Z'$ masses of a few TeV, as the LHC lower limits with approximately 5~fb$^{-1}$ of data fall in the vicinity of $2-2.5$~TeV for the 
considered models~\cite{cms612}.

In Section~\ref{DMZ'} we study the phenomenology of a $Z'$ scenario in which SM particles are uncharged under the new $U(1)'$. In this case, SM particles interact with a new sector through kinetic and mass mixing of the $Z'$ with the $Z$.  Such a $Z'$ could act as a dark matter mediator, with isospin-violating dark matter scattering arising naturally.  Then, there are two complementary ways to test such a model: the production of $Z'$ resonances in collider experiments, and the direct detection of dark matter particles. We combine the data from LHC $Z'$ searches~\cite{Collaboration:2011dca, Timciuc:2011ji} and the XENON100 dark matter experiment~\cite{Aprile:2012nq} with global electroweak data to constrain the kinetic and mass mixing angles. We find that the electroweak, collider, and dark matter data provide comparable limits.

\section{\label{detection} $Z'$ bosons at the LHC}

If the $Z'$ couples to standard model quarks and leptons, it could be detected at the LHC as a resonance in the dilepton channel through the experimentally well-studied process \mbox{$pp\rightarrow Z' \rightarrow l^+l^-$}~\cite{Erler:2011ud, Erler:2011iw, Cvetic:1995zs, Rizzo:2006nw, Petriello:2008zr, Dittmar:2003ir}. We focus on the dielectron and dilmuon channels, though decays to $\tau$-lepton pairs can also be useful~\cite{cmstaulep}. The SM background to this process is fairly small, consisting mostly of Drell-Yan (DY) $Z/\gamma^*$ events and a smaller number of $t \bar{t}$ and multijet events~\cite{Collaboration:2011dca, Timciuc:2011ji}.

The differential cross section for the DY process is~\cite{Carena:2004xs}
\begin{equation}
\frac{d\sigma}{dQ^2}= \frac{1}{s}\sigma(Z' \rightarrow l^+l^-) \, W_{Z'}(s,Q^2) + \text{interference terms}.
\label{xsec_interference}
\end{equation}
The first term on the right hand side is the pure $Z'$ contribution, factored into two parts:  a hadronic structure function containing the QCD dependence, $W_{Z'}(s,Q^2)$, and the partonic cross section,
\begin{equation}
\sigma(Z' \rightarrow l^+l^-) = \frac{1}{4\pi}\frac{g_{l_L}^2+g_{l_R}^2}{288}\frac{Q^2}{(Q^2-M_{Z'}^2)^2+M_{Z'}^2\Gamma_{Z'}^2}\, .
\end{equation}
The interference terms with photons and the SM $Z$ are given in the Appendix. For narrow resonances, which we define as $\Gamma_{Z'} < 0.1 M_{Z'}$, interference effects can be neglected. This is an excellent approximation near the resonance peak. In the region slightly off peak, the interference terms can significantly alter the shape of the invariant mass distribution~\cite{Carena:2004xs, Accomando:2010fz}.  

The partial width for decay into a massless fermion pair $f\bar{f}$ is given by
\begin{equation}
\Gamma_{Z'}^{f} = \frac{M_{Z'}}{24\pi}(g_{f_L}^2+g_{f_R}^2)\, ,
\end{equation}
where the $g_{f_{L,R}}$ are the fermion couplings, which can be written in terms of the overall $Z'$ coupling and the fermion charges $z_f$:
\begin{equation}
g_{f_{L,R}}^2=g_{Z'}^2 \, z_{f_{L,R}}^2=\frac{4\pi\alpha}{\cos^2{\theta_W}} \, z_{f_{L,R}}^2 \, .
\end{equation}
We take the masses of the quarks to be negligible compared to $M_{Z'}/2$.


We assume for simplicity that the $Z'$ decays only to SM fermions. However, in models with extra fermions, the $Z'$ might also have non-SM decays. Thus, the total width is generally a free parameter. If the decay rate to new fermions is large, the $Z'$ mass limits could be significantly relaxed due to the reduced SM branching fractions~\cite{Chang:2011be}.

Dilepton resonances are not the only viable channel for $Z'$ detection.  Past work has considered detection using decays to top quarks~\cite{Barger:2006hm, delAguila:2009gz, Abazov:2011gv, Chiang:2011kq} and third generation fermions~\cite{Godfrey:2008vf, Diener:2009vq, Diener:2010sy}, weak boson pair production~\cite{Eboli:2011ye}, and weak charge measurements in atomic parity violation experiments~\cite{Diener:2011jt, Barger:1987xw}.  The latter channels are particularly useful in the case of leptophobic or non-universal $Z'$ models~\cite{Barger:1996kr}.

Other new physics, including Randall-Sundrum gravitons or sneutrinos,  could also be detected via a dilepton resonance similar to a $Z'$. There have been several discussions of how to differentiate such resonances from a $Z'$~\cite{Chiang:2011kq, Osland:2009tn}.

\section{\label{Sim} Simulation}

We have analyzed the characteristic features of $Z'$ production at the LHC using an expanded version of the simulation code FEWZ~\cite{Gavin:2012kw, Gavin:2010az} (see Appendix). All calculations are done to NLO or NNLO in QCD using the MSTW2008 PDF sets~\cite{Martin:2009iq}. Factorization and renormalization scales are set to $\mu_F = \mu_R = M_{Z'}$~\cite{Gavin:2010az}. 

We adopt the following standard acceptance cuts in our analysis:
\begin{align*}
p_T^\mathit{l} &> 20~\rm{GeV}\,, \\
|\eta_l| &< 2.5\,, \\
|y_{Z'}| &>0.8 \text{\ \  (for $A_{FB}$ only)\,.}
\end{align*}
The first two of these cuts are the same as those applied to the muon channel by CMS and ATLAS~\cite{Timciuc:2011ji, Collaboration:2011dca}. The third is used to define a forward-backward asymmetry. The intial quark direction cannot be measured directly at proton-proton colliders.  However, the boost direction of the $Z'$ is preferentially in the direction of the quark, not the antiquark, especially for large dilepton rapidity 
$|y_{ll}|$. Valence quarks are much more likely than any other partons to carry a large fraction of the proton momentum. Thus, if the dilepton rapidity is large, the boost is preferentially in the valence quark direction. However, for small dilepton rapidities, the initial momenta of the quark and antiquark have similar magnitudes, so we cannot use the parton distributions to distinguish between them. Therefore, placing a cut on the rapidity of the final dilepton system allows for a measurement of $A_{FB}$~\cite{Dittmar:2003ir}.

We consider several common benchmark $Z'$ models. The first, the sequential standard model (SSM), has couplings identical to those of the SM $Z$. It is a common benchmark in experimental searches. Another theoretically interesting case is the $Z'_{B-L}$, noteworthy because $U(1)_{B-L}$ satisfies anomaly cancellation conditions without the presence of exotic fermions or non-universal couplings ~\cite{Carena:2004xs,Barger:2008wn, FileviezPerez:2012mj} and may be a remnant of string theory~\cite{Braun:2005ux, Ovrut:2012wg, Anchordoqui:2012wt}.

A theoretically well-motivated class of $Z'$ models derives from breaking the $E_6$ gauge group via the chain~\cite{Hewett:1988xc}
\begin{equation}
\label{breaking chain}
SU(5) \times U(1)_{\psi} \times U(1)_{\chi} \rightarrow 
SU(3)_c \times SU(2)_L \times U(1)_Y \times U(1)_{\psi} \times U(1)_{\chi}\,. \nonumber
\end{equation}
The new $U(1)$ factors are associated with two neutral gauge bosons, $Z_\psi$ and $Z_\chi$. After symmetry breaking, they mix to form the mass eigenstates $Z'$ and $Z'',$ with the mixing parameterized by an angle $\beta$~\cite{Barger:1986nn}: 
\begin{align}
Z' &= Z_{\chi} \sin{ \beta}+Z_{\psi} \cos{ \beta}\,,  \nonumber \\ 
Z'' &= Z_{\chi} \cos{ \beta}-Z_{\psi} \sin{ \beta}\,.
\label{beta}
\end{align}
Since the $Z'$ and $Z''$ are assumed to be heavy compared to the SM $Z$-boson,  any mixing with the $Z$ is negligible and ignored here. For now, we only consider the lower-mass $Z'$ for the  three specific cases $Z'_\psi$ for $\beta = 0$, $Z'_\chi$ for $\beta = \pi/2$, and $Z'_\eta$ for $\tan{\beta}=\sqrt{3/5}$.  We explore $Z'$ and $Z''$ detection in this class of models in Section~\ref{E6 analysis}.

Finally, we consider a left-right symmetric model, $Z'_{LR}$~\cite{mohapatra2003unification}.  The $Z'$ couples to the current 
\begin{equation}
J^\mu_{LR}=\alpha_{LR} J^\mu_{3R}-\frac{1}{2\alpha_{LR}}J^\mu_{B-L}\, ,
\end{equation}
where $\alpha_{LR} = \sqrt{\cos^2{\theta_W} g^2_R/\sin^2{\theta_W} g^2_L -1}$ and $\sin^2{\theta_W} = 0.22255$~\cite{Nakamura:2010zzi}. We study an example $Z'_{LR}$ with $g^2_R=g^2_L$ {\it i.e.,} $\alpha_{LR}\approx1.58$~\cite{Dittmar:2003ir}.

Table \ref{couplings_table} summarizes the $Z'$ couplings to SM fermions for these models. The overall gauge coupling strength is a free parameter, but it is often chosen to be consistent with a grand unification scenario. We follow this approach and set $g_{Z'}^2 = \frac{4\pi\alpha}{\cos^2{\theta_W}}$~\cite{Petriello:2008zr}. (This factor is not included in Table \ref{couplings_table}.)

\begin{table*}
\begin{center}
{\renewcommand{\arraystretch}{1.75}
\begin{tabular} {| c| c | c | c | c  | c| c| c|}
\hline 
 & $z^u_L$ & $z^d_L$ &$z^e_L$ & $z^u_R$ & $z^d_R$ &$z^e_R$ & $z^\nu_L$  \\ \hline\hline

$\psi$ & $\frac{\sqrt{10}}{12}$ &$\frac{\sqrt{10}}{12}$  & $\frac{\sqrt{10}}{12}$ & $\frac{-\sqrt{10}}{12}$ & $\frac{-\sqrt{10}}{12}$ & $\frac{-\sqrt{10}}{12}$ &$\frac{\sqrt{10}}{12}$ \\

$\chi$ & $\frac{-1}{2\sqrt{6}}$& $\frac{-1}{2\sqrt{6}}$&$\frac{3}{2\sqrt{6}}$ & $\frac{1}{2\sqrt{6}}$&$\frac{-3}{2\sqrt{6}}$ & $\frac{1}{2\sqrt{6}}$ & $\frac{3}{2\sqrt{6}}$\\

$\eta$ &$\frac{1}{3}$ &$\frac{1}{3}$ & $\frac{-1}{6}$&$\frac{-1}{3}$ &$\frac{1}{6}$ & $\frac{-1}{3}$&$\frac{-1}{6}$\\

$B-L$ & $\frac{\sqrt{5}}{6\sqrt{2}}$ &$\frac{\sqrt{5}}{6\sqrt{2}}$ & $-\frac{\sqrt{5}}{2\sqrt{2}}$& $\frac{\sqrt{5}}{6\sqrt{2}}$& $\frac{\sqrt{5}}{6\sqrt{2}}$& $-\frac{\sqrt{5}}{2\sqrt{2}}$&$-\frac{\sqrt{5}}{2\sqrt{2}}$\\

LR & $-\frac{1}{6 \alpha_{LR}}$&$-\frac{1}{6 \alpha_{LR}} $& $\frac{1}{2 \alpha_{LR}}$& $-\frac{1}{6 \alpha_{LR}}+\frac{ \alpha_{LR}}{2}$&$-\frac{1}{6 \alpha_{LR}}-\frac{ \alpha_{LR}}{2} $& $\frac{1}{2 \alpha_{LR}}-\frac{ \alpha_{LR}}{2}$ &$\frac{1}{2 \alpha_{LR}}$\\ \hline 
\end{tabular}}
\end{center}
\caption{\label{couplings_table} Z$'$ gauge charges. For the LR model, $\alpha_{LR}\approx 1.58$}
\end{table*}

In Table \ref{widths} we list for each model the total width, normalized by $M_{Z'}$, as well as the branching fractions to leptons, quarks, and neutrinos, assuming no decays to non-SM particles. For our calculations, we use $\alpha^{-1}(M_{Z}) = 128$.

\begin{table*}
\begin{center}
\begin{tabular} {| c | c | c | c | c | c |}
\hline 
Model & $\Gamma/M_{Z'}$ & BF($\mu^+\mu^-$)  & BF($t\bar{t}) $ &BF(hadrons) & BF($\nu\bar{\nu}$) \\ \hline \hline
$\psi$ & 0.005 & 0.04& 0.12 &0.80 & 0.07\\
$\chi$ & 0.012& 0.06& 0.03&0.65& 0.16\\
$\eta$ & 0.006 & 0.04& 0.16&0.87& 0.02\\
$B-L$ &  0.014 & 0.15 & 0.05&0.31& 0.23\\
LR & 0.022 & 0.03& 0.10&0.90& 0.02\\
SSM & 0.026 & 0.03& 0.10&0.73& 0.18\\
\hline 
\end{tabular}
\end{center}
\caption{\label{widths}$Z'$ decay widths and branching fractions, assuming no non-SM fermion decays.}
\end{table*}

\begin{figure}
\begin{center}
\includegraphics[width=3in]{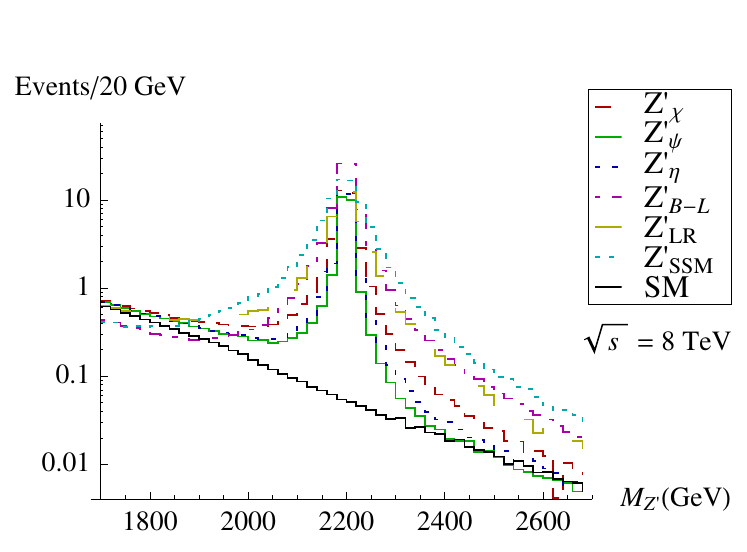}
\includegraphics[width=3in]{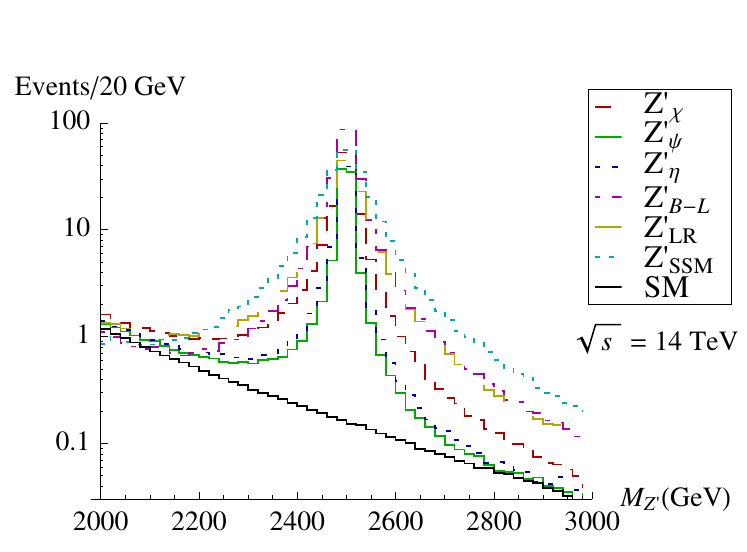} 
\end{center}
\caption{\label{BW peak}Simulated NNLO dimuon invariant mass spectrum for a variety of $Z'$ models with $M_{Z'} = 2.2$~TeV and $\sqrt{s} = 8$~TeV (left) and $M_{Z'} = 2.5$~TeV and $\sqrt{s} = 14$~TeV (right), both with a luminosity of ${\cal L}=100$~fb$^{-1}$.  } 
\end{figure}

In Fig.~\ref{BW peak} we show the shape of the dilepton mass spectrum for a variety of $Z'$ models with 
$M_{Z'} = 2.2$~TeV at $\sqrt{s} = 8$~TeV (LHC8), and $M_{Z'} = 2.5$~TeV at $\sqrt{s} =14$~TeV (LHC14).  The SM background is very small in the resonance mass range, so for luminosities of 20~fb$^{-1}$  for $\sqrt{s} = 8$~TeV or 100~fb$^{-1}$ for 
$\sqrt{s} = 14$~TeV, these resonance peaks should be clearly distinguishable.  Distributions are calculated at next-to-next-to-leading order. We neglect detector effects such as energy resolution smearing. For ATLAS, the resolution width is about $1\%$ for electrons and $5\%$ for muons~\cite{Collaboration:2011dca}. For CMS, the resolutions widths are $1-2\%$ for electrons and $4-7\%$ for muons~\cite{Timciuc:2011ji}. As can be seen in Table \ref{widths}, the resolutions are generally comparable to or larger than the decay widths for the $Z'$ (assuming that non-SM decay rates are not large). Therefore a precise measurement of the $Z'$ width will likely be difficult \cite{Accomando:2010fz}.  As we show later, there are observables that do not depend strongly on the decay width but still provide useful information about the properties of the $Z'$.

\begin{figure}
\includegraphics[width=3in]{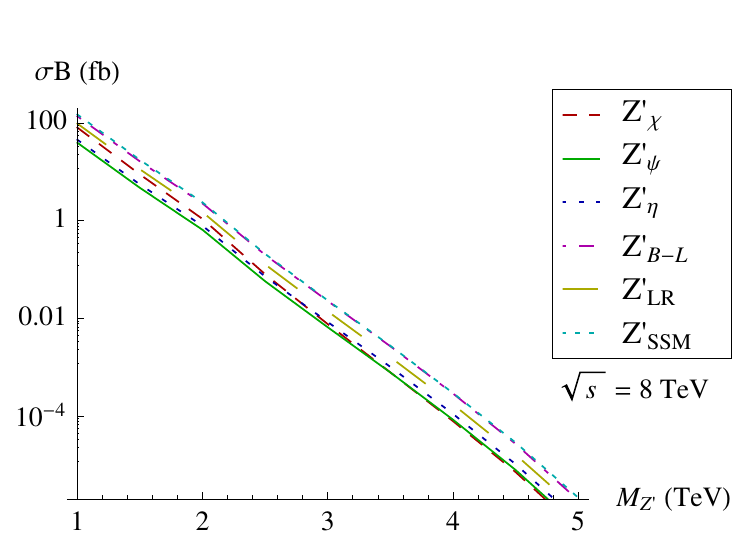}
\includegraphics[width=3in]{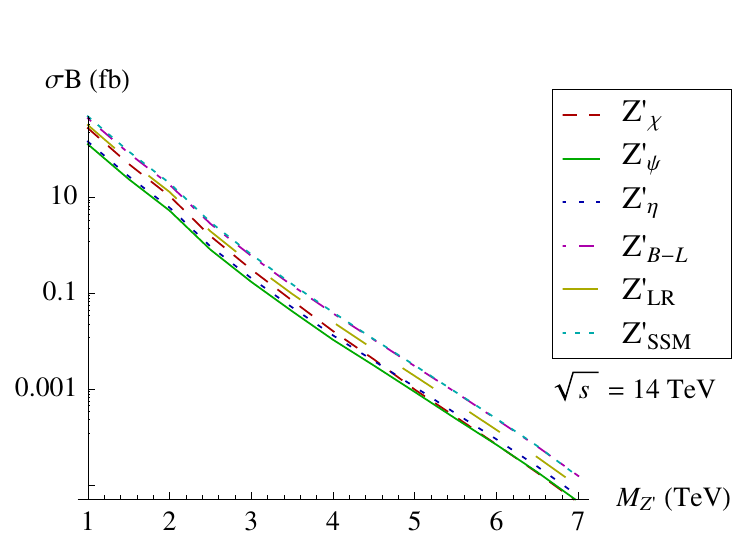}
\caption{\label{mass dependence}NLO cross section for the process $Z' \rightarrow l^+l^-$ versus Z$'$ mass, integrated over the dilepton invariant mass peak region ($\pm 3 \Gamma$) for $\sqrt{s} = 8$~TeV (left) and $\sqrt{s} = 14$~TeV (right). }
\end{figure}

By integrating over the peak region, we can determine the cross section as a function of $M_{Z'}$ for each model. In Fig.~\ref{mass dependence}, we show the mass dependence of the integrated peak ($\pm 3 \Gamma_{Z'}$) cross section for our six model examples at NLO. Note that the mass dependence is largely model-independent. 

\section{\label{fits} Analysis}

We now present a few empirical formulas from our simulations, which allow the cross section and the differential distributions over a wide range of model parameters to be predicted.  They are also useful in extracting coupling information from experimental data. 

The LHC production cross section of the $Z'$ depends on the mass, width, and couplings. Under the assumption of generational universality, this dependence can be parameterized in the narrow width approximation by
\begin{align}
\label{fit}
 \sigma(&pp \rightarrow Z' \rightarrow l^+l^-) =\small{\left[ p\, (g_{u_L}^2+g_{u_R}^2)+ (1-p) \,(g_{d_L}^2+g_{d_R}^2) \right] \text{B}(Z' \rightarrow l^+l^-)  \,f(r_{Z'})} \, . 
\end{align}
The parameter $p$ is model-independent but varies with $r_{Z'}$. It quantifies the fractional contribution to the cross section from up-type quark events. The dependence on the $Z'$ mass is contained in $f(r_{Z'})$.  We use the empirical representation~\cite{Leike:1998wr},
\begin{equation}
\label{f(M)}
f(r_{Z'}) = \sigma_0 r_{Z'}^a \left( \frac{1}{r_{Z'}}-1\right)^b\,,
\end{equation}  
of the structure function dependence, where $r_{Z'} = \frac{M_{Z'}}{\sqrt{s}}$.

As in Refs.~\cite{Petriello:2008zr, Carena:2004xs}, we define the quantities
\begin{align}
\label{def_c}
c_q &= \frac{M_{Z}'}{24 \pi \Gamma_{Z'}}(g_{q_L}^2+g_{q_R}^2) (g_{e_L}^2+g_{e_R}^2) = (g_{q_L}^2+g_{q_R}^2) \text{ B}(Z' \rightarrow l^+l^-)\,, \\
e_q &= \frac{M_{Z}'}{24 \pi \Gamma_{Z'}}(g_{q_L}^2-g_{q_R}^2) (g_{e_L}^2-g_{e_R}^2)\,.
\end{align}
In this notation, the cross section formula takes the simple form:
\begin{equation}
\label{peak xsec}
 \sigma(pp \rightarrow Z' \rightarrow l^+l^-) = \left[p \, c_u + (1-p) \, c_d\right] f(r_{Z'}) \, .
 \end{equation}

\begin{figure}
\includegraphics[width=3in]{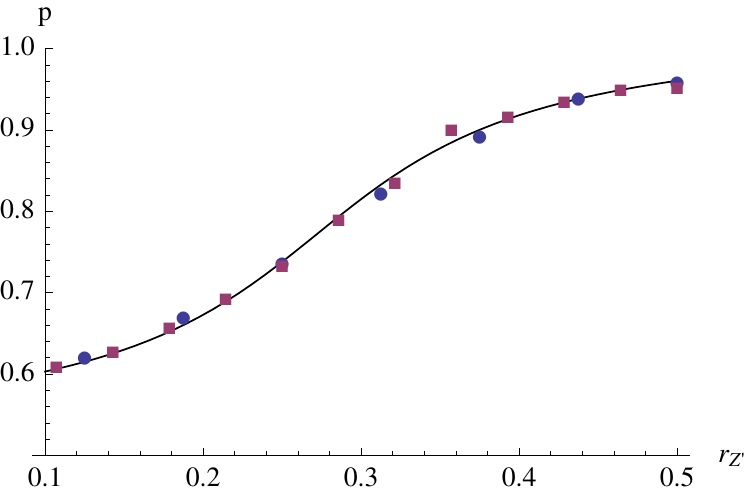}
\caption{\label{p_fit}The parameter $p$ (as defined in Eq.~\ref{fit}) as a function of $r_{Z'} = \frac{M_{Z'}}{\sqrt{s}}$, along with an approximate fit to an arctangent function. Notice that simulated values (points) are shown for both 8~TeV (squares) and  14~TeV (circles) and agree well.}
\end{figure}

We use a NLO FEWZ simulation of the six model examples in fits of the parameters of Eqs.~(\ref{fit}) and~(\ref{f(M)}). First, we fix $r_{Z'}$ and fit the simulated cross sections for the six models to the form $A c_u + B c_d$, then take $p = \frac{A}{A+B}$. We repeat this procedure for several values of $r_{Z'}$  over the range, $0.1-0.6$ ($M_{Z'} = 1.0-4.5$~TeV for 
$\sqrt{s} = 8$~TeV; $M_{Z'} = 2.0-7.0$~TeV for $\sqrt{s} = 14$~TeV), to determine how $p$ changes with $r_{Z'}$.   We find that this dependence can be approximated by the function 
\begin{equation}
p(r_{Z'}) = 0.77 - 0.17 \tan^{-1}{(2.6 - 9.5\, r_{Z'})}\,.
\end{equation}
The result is shown in Fig.~\ref{p_fit}.

\begin{figure}
\includegraphics[width=3in]{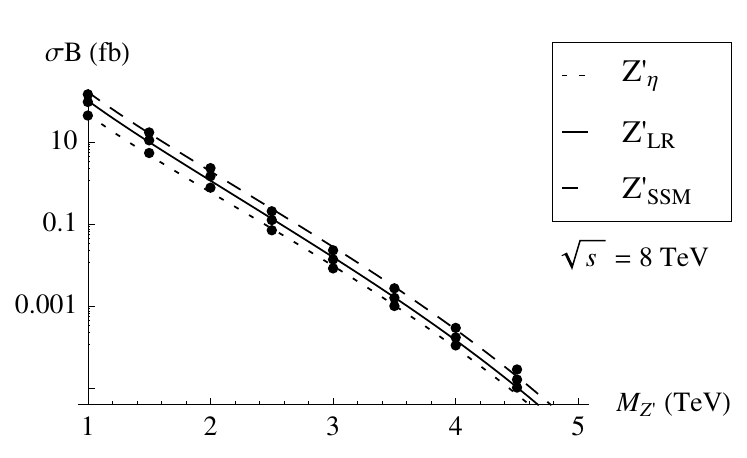}
\includegraphics[width=3in]{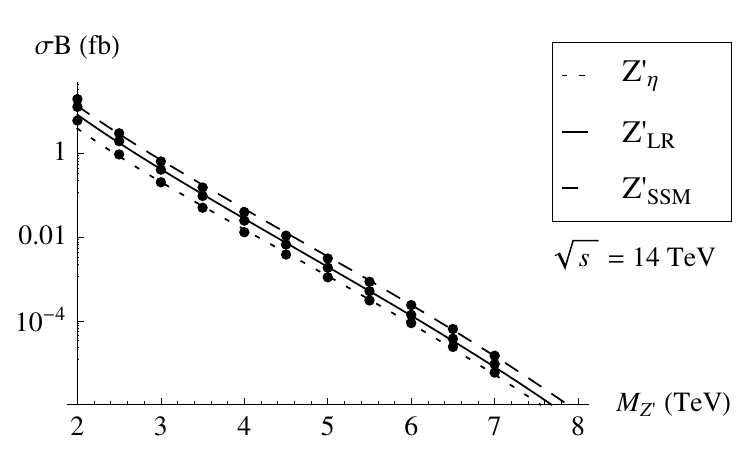}
\caption{\label{mass fit} FEWZ simulation (points) and fitted curve (lines) of the integrated $Z'$ cross section for $\sqrt{s} = 8$~TeV (left) and $\sqrt{s} = 14$~TeV (right).  We show only the $E_6$ $\eta$, left-right, and sequential models for simplicity.}
\end{figure}

To determine $f(r_{Z'})$, we fit the normalization $A+B$ to the form given in Eq.~(\ref{f(M)}). Our best fit values are 
\begin{equation}
f(r_{Z'}) = \small{\left\{ 
\begin{array}{l l}
 (3200\,{\rm fb}) \ r_{Z'}^{15.0} \left( \frac{1}{r_{Z'}}-1\right)^{17.5}  & \quad \mbox{$\sqrt{s}$ = 8 TeV}\,,\\
(43.3\,{\rm fb}) \ r_{Z'}^{13.1} \left( \frac{1}{r_{Z'}}-1\right)^{16.8}  & \quad \mbox{$\sqrt{s}$ = 14 TeV}\,.\\ \end{array} \right. }
\end{equation}
A comparison of our fit to the simulation is plotted in Fig.\  \ref{mass fit}.  This fit captures the dependence on $r_{Z'}$ within $20\%$ over the entire range $M_{Z'} = 1-7$~TeV.   

After a $Z'$ is detected, the next step will be to measure its couplings.  Measuring couplings with precision requires a large number of events, which may not be the situation for a very heavy or weakly coupled $Z'$. However, with enough events, an accurate measurement of $c_q$ and $e_q$ is possible, as we demonstrate by considering a set of simulated measurements of a 2.5~TeV $Z'$ at $\sqrt{s} = 14$~TeV.

The coupling combinations $c_q$ and $e_q$ of Eq.~(\ref{def_c}) can be determined by considering the differential cross section, integrated over the resonance peak:
\begin{align}
\label{diff xsec}
\begin{split}
\frac{d\sigma_{int}}{dy \,d\cos{\theta}}  = &\int^{M_{Z'}+3\Gamma}_{M_{Z'}-3\Gamma} \frac{d\sigma}{dy \,d\cos{\theta}\, dQ} dQ \\
= & \ 3/8 \,(1+\cos^2{\theta})\,[p \,c_u\, h_1^u(y) + (1-p)\, c_d \,h_1^d(y)] \,f(r_{Z'}) \\
& + \cos{\theta}\,[p \,e_u\, h_2^u(y) + (1-p)\, e_d \,h_2^d(y)] \,g(r_{Z'} )\,.
\end{split}
\end{align}
The functions $h_{1,2}^q$ are normalized so that integrating over $\cos{\theta}$ and $y$ yields Eq.~(\ref{peak xsec}). $g(r_{Z'} )$ represents the mass dependence of the $\cos{\theta}$ term, similar to $f(r_{Z'} )$. For $M_{Z'} = 2.5$~TeV, its value is $g(r_{Z'} ) = 506$~fb. For the same mass, we find $f(r_{Z'} ) = 1050$~fb.

\begin{figure}
\includegraphics[width=3in]{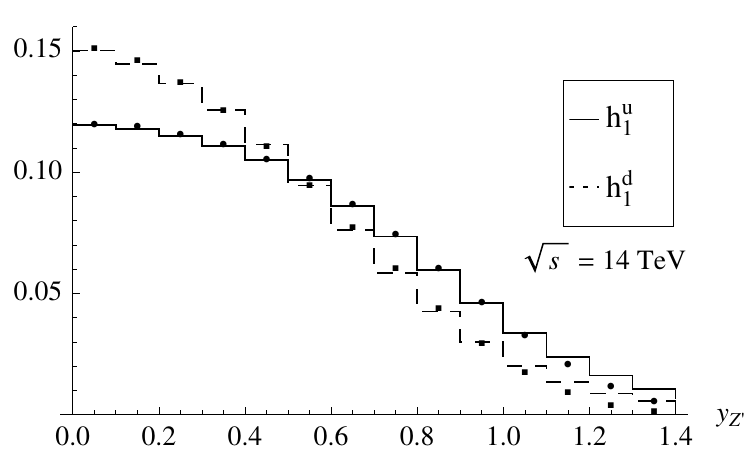}
\includegraphics[width=3in]{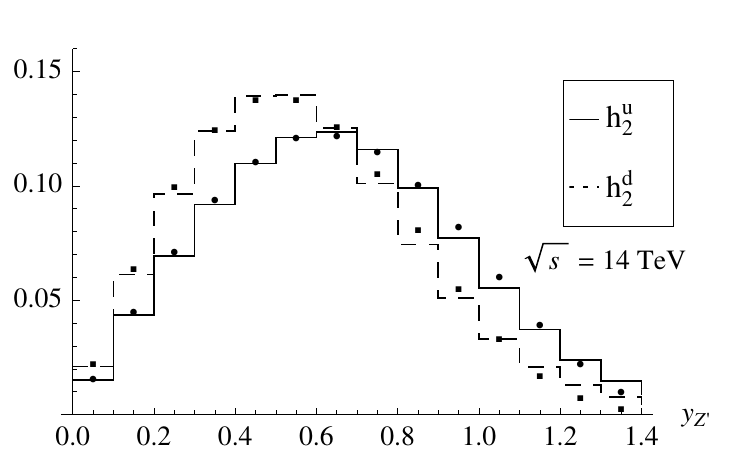} \\
\caption{\label{y_up_down}Normalized distributions for $\frac{d\sigma}{dy}$ (left) and $\frac{d(F-B)}{dy}$ (right)  for a $Z'$ that couples to only up-type (solid) or down-type (dashed) quarks. We have set $M_{Z'} = 2.5$~TeV at LHC14. Points are simulated data and lines are fitted curves.}
\end{figure}

We have already determined $p$, $f(r_{Z'})$, and $g(r_{Z'})$,  so all that remains is to find $h_{1,2}^q$. In order to separate the up and down contributions to the differential cross section, we define two distinct scenarios in which the $Z'$ couples exclusively to one type of quark (and to leptons). Then, for each scenario we simulate two differential distributions. The first is the dilepton rapidity distribution, which allows us to determine $h_1^q$. To determine $h_2^q$, instead of using the total cross section, we consider the quantity $\frac{d(F-B)}{dy}$, where $F$ is the number of lepton pairs scattered in the forward ($\cos \theta_{Z'} > 0$) direction in the Collins-Soper frame~\cite{PhysRevD.16.2219} and $B$ is the number scattered in the backward direction. 
The resulting distributions are easily distinguishable, as can be seen in Fig.~\ref{y_up_down}. In addition to the simulated NLO data, 
Fig.~\ref{y_up_down} shows approximate curves $h_{1,2}^q(y)$. The specific curves used to approximate the normalized distributions at 14~TeV are
\begin{align}
h_1^u(y) &= \frac{0.59}{e^{4.6 (y-0.84)}+1}\,, \nonumber \\
h_1^d(y) &= \frac{0.78}{e^{4.6 (y-0.63)}+1}\,, \nonumber \\
h_2^u(y) &= \frac{2.5 \,(1-e^{-0.60 y})}{e^{5.7 (y-0.84)}+1}\,, \nonumber \\
h_2^d(y) &= \frac{11\, (1-e^{-0.20 y})}{e^{5.8 (y-0.63)}+1}\,.
\end{align} 

By fitting observed data to Eq.~(\ref{diff xsec}), one can determine the four coefficients $c_q$ and $e_q$. To demonstrate the feasibility of this method and estimate the statistical error, we use our simulation as a pseudo-experiment.  For each reference model in Section~\ref{Sim}, we generate binned distributions for $\frac{dN}{dy}$ and $\frac{d(F-B)}{dy}$. To extract  $c_q$ and $e_q$, we fit these distributions to linear combinations of $h^u_{1,2}$ and $h^d_{1,2}$ by minimizing $\chi^2$. To determine the boundaries of the confidence regions, we vary $c_q$ and $e_q$ and calculate the $\chi^2$ value at each point for an ``average'' experiment using the method described in Appendix A of \cite{deGouvea:1999wg}.

Figures~\ref{error} and~\ref{errora}  show the 95\% confidence level (C.L.) regions for our example case of a 2.5~TeV $Z'$ at 14~TeV. We see that 100~fb$^{-1}$ allows for some model differentiation, while for 1~ab$^{-1}$ of luminosity the confidence regions are narrow.  In the right panel of Fig.~\ref{error}, the red dashed contour shows the values of $c_u$ and $c_d$ for the $E_6$ models as a function of the mixing angle $\beta$, from which we can see that some model differentiation should be possible. The tilt of the ellipses arises from the requirement that the up and down quark contributions are summed to give the total cross section, restricting $c_u$ and $c_d$ to lie on a line. 

\begin{figure}
\includegraphics[width=3in]{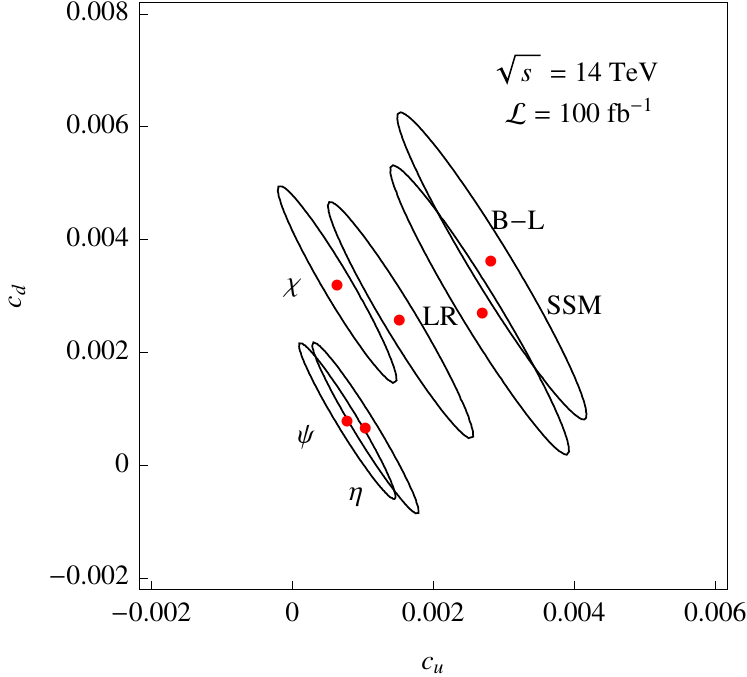}
\includegraphics[width=3in]{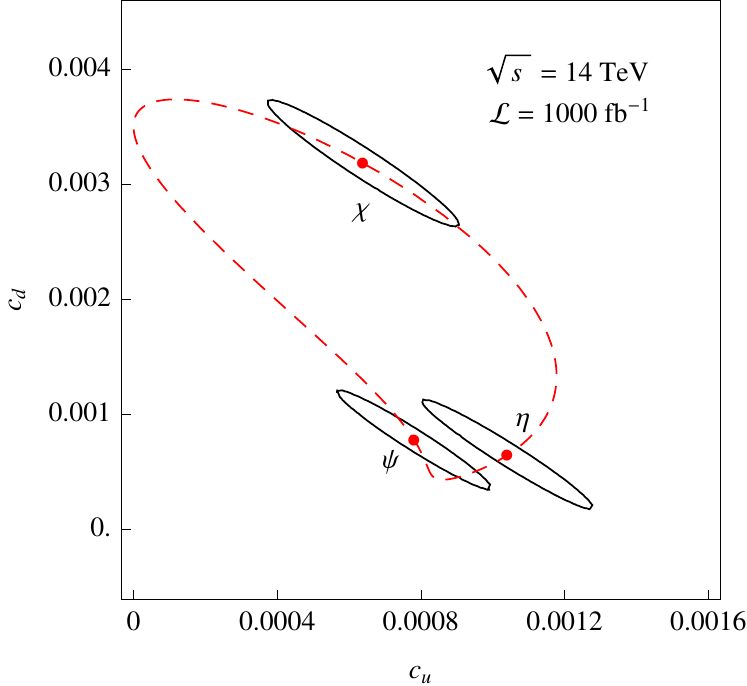} \\
\caption{\label{error}95\% C.L. regions for the couplings $c_u$ and $c_d$ for an average experiment (statistical errors only) for ${\cal L} = 100$~fb$^{-1}$ (left) and ${\cal L} = 1$~ab$^{-1}$ (right). The values of $c_u$ and $c_d$ for the $E_6$ family lie on the dashed red contour. Points are the theoretical values. }
\end{figure}

\begin{figure}
\includegraphics[width=3in]{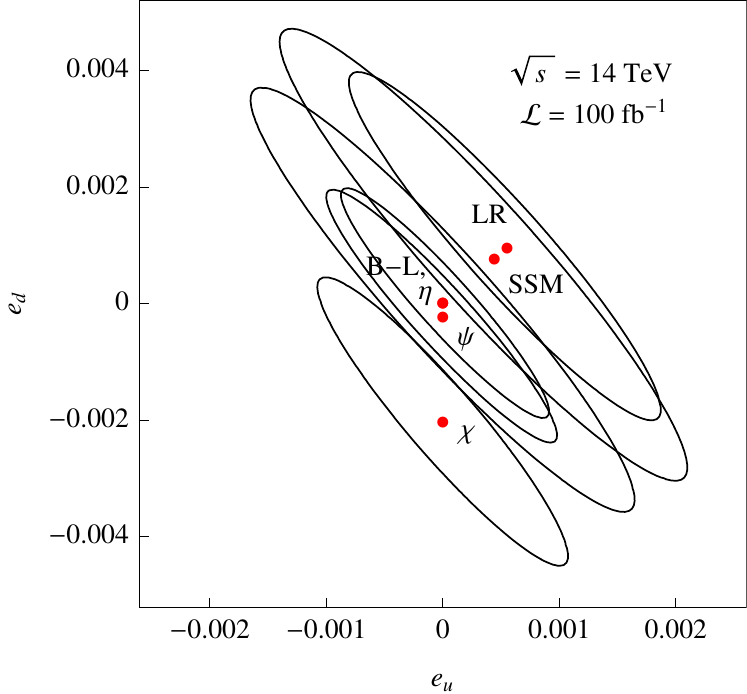}
\includegraphics[width=3in]{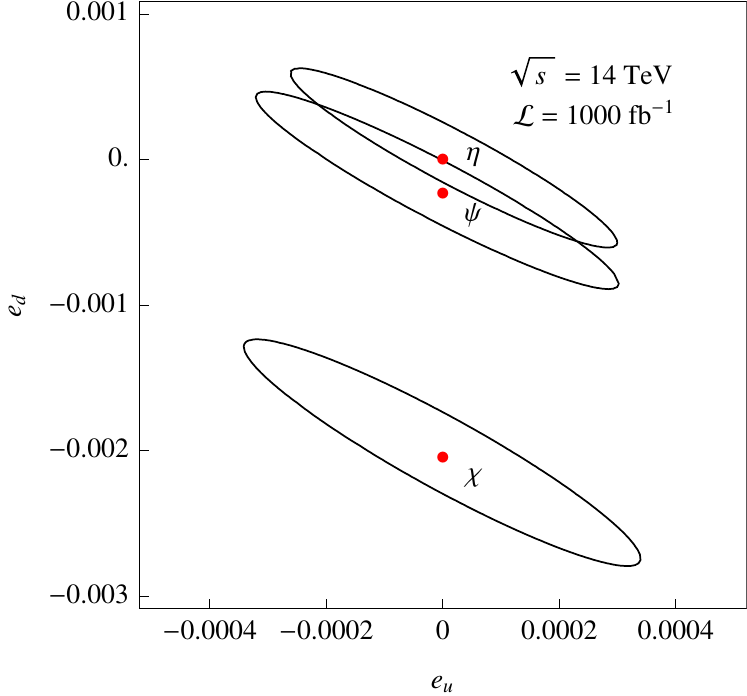}
\caption{\label{errora}95\% confidence level regions for the couplings $e_u$ and $e_d$ for an average experiment for 
${\cal L} = 100$~fb$^{-1}$ (left) and ${\cal L} = 1$~ab$^{-1}$ (right). Points are the theoretical values. }
\end{figure}

Within the $E_6$ model class, our fit can also be used to place limits on $\beta$ using the least squares method. Using Eqs.~(\ref{beta}) and~(\ref{def_c}), we can write Eq.~(\ref{diff xsec}) in terms of $\beta$ instead of $c_q$ and $e_q$. At the 1-sigma level, we find
\begin{align}
0.44 \pi &< \beta_\chi < 0.54  \pi\,, \nonumber \\
0.97  \pi &< \beta_\psi < 1.03  \pi  \text{ or } 1.14  \pi < \beta_\psi < 1.17  \pi\,,  \\
0.20  \pi &< \beta_\eta < 0.22  \pi\,. \nonumber 
\end{align}

\begin{figure}[!]
\includegraphics[width=3in]{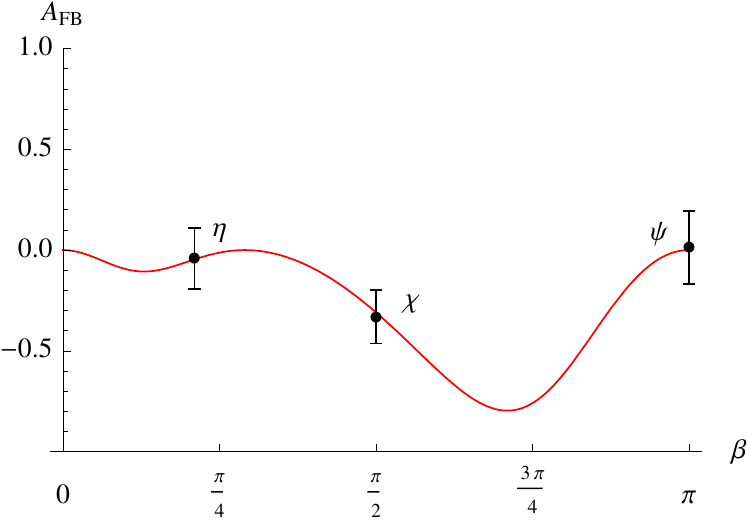}
\includegraphics[width=3in]{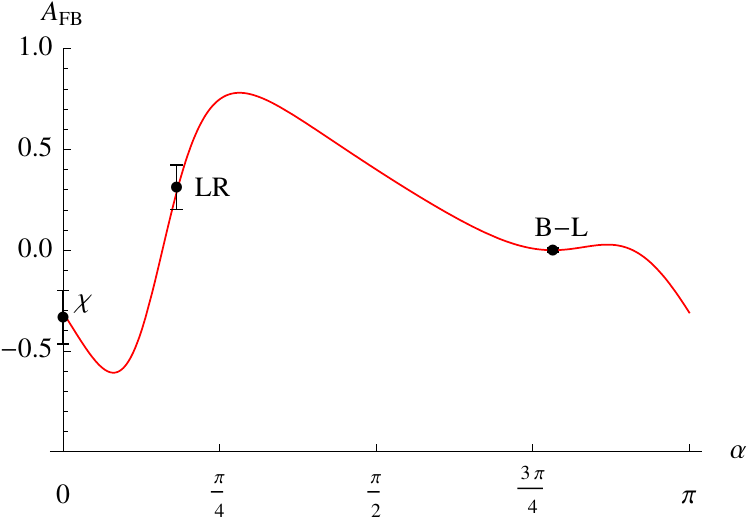}
\caption{\label{afb_errorbands}Predicted on-peak forward-backward asymmetry versus mixing angle $\beta$ for the $E_6$ models (left) and $\alpha$ for the LR models (right), along with FEWZ simulated values and statistical errors for $M_{Z'}=2.5$~TeV, $\sqrt{s} = 14$~TeV, and $\mathcal{L} = 100~\text{fb}^{-1}$.} 
\end{figure}

If the number of events is low, an analysis can still be done by integrating over the distributions $h_{1,2}^q$, leading to a system of four equations in four unknowns that can be inverted~\cite{Petriello:2008zr}.   Note that we have included only statistical errors. For information on the effect of PDF errors, see Ref.~\cite{Petriello:2008zr}.

From our fit, we can also determine the forward-backward asymmetry, 
\begin{equation}
A_{FB} = \frac{F-B}{F+B}\,.
\end{equation}
On the $Z'$ peak, $A_{FB}$ depends solely on the $Z'$ couplings to fermions.  Using Eq.~(\ref{diff xsec}), integrated over appropriate ranges of $\cos{\theta}$, we can write
\begin{align}
\label{AFB}
A_{FB} &= \frac{(p \,a_2^u\, e_u +(1-p) \,a_2^d \,e_d)}{(p \,a_1^u \,c_u + (1-p)\, a_1^d \,c_d)}\frac{g(M_{Z'} ) }{f(M_{Z'} ) }\,,
\end{align}
where $a_{1,2}^q = 2 \int^{y_{max}}_{y_{min}} h_{1,2}^q(y) \,dy$. We choose $y_{max} = 2.5 $ and $y_{min} = 0.8$, the rapidity cuts discussed in Section~\ref{Sim}.

 In the left panel of Fig.~\ref{afb_errorbands}, we show the simulated values of $A_{FB}$ along with statistical uncertainties for three $E_6$ models with 100~fb$^{-1}$ of data. The red curve shows the predicted values for $A_{FB}$ versus the mixing angle $\beta$ defined in Eq.~(\ref{beta}). The couplings of the $LR$, $B-L$, and $\chi$ models can also be parameterized (up to a normalization factor) by an angle $\alpha$ with~\cite{Erler:2011ud}
\begin{equation}
\label{lr models}
Z' = \cos{\alpha} \, Z_{\chi}+\sin{\alpha} \,Z_{Y}\,.
\end{equation}
This is equivalent to 
\begin{equation}
Z'_{LR} = \cos{\theta_{LR}} (-Z_{B-L})+\sin{\theta_{LR}} \,Z_{R}\,.
\end{equation}
where $\arctan({\alpha_{LR}}) = \theta_{LR} = \alpha + \arctan{\sqrt{2/3}}$. As for the three $E_6$ models, we show $A_{FB}$ versus $\alpha$ in the right panel of Fig~\ref{afb_errorbands}.

\begin{figure}[!]
\includegraphics[width=3in]{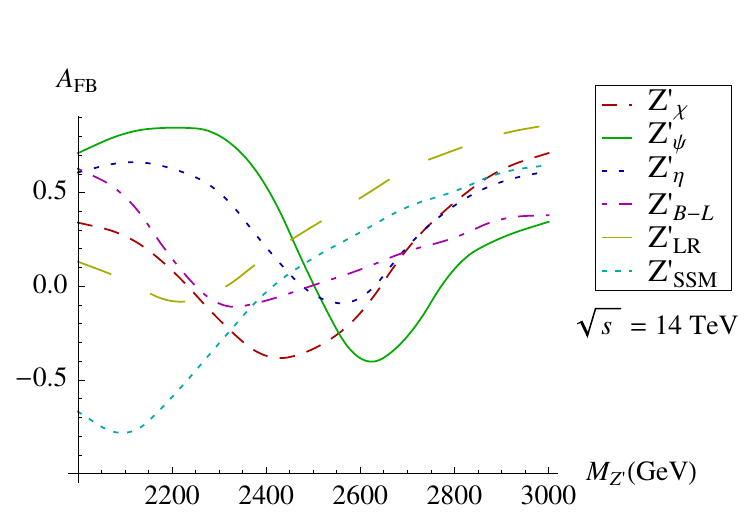}
\caption{\label{afb}Simulated forward-backward asymmetry for a number of $Z'$ models with $M_{Z'}=2.5$~TeV
 and $\sqrt{s} = 14$~TeV.} 
\end{figure}

Off peak, the $Z'$ interference with the SM background contributes to the asymmetry, so $A_{FB}$ varies significantly with dilepton mass.  
Figure~\ref{afb} shows the effect of this interference on $A_{FB}$.  Both the shape of the curve and the peak value are highly model-dependent. 

\section{\label{E6 analysis} $E_6$ Models}
\begin{figure}[!t]
\includegraphics[width=3in]{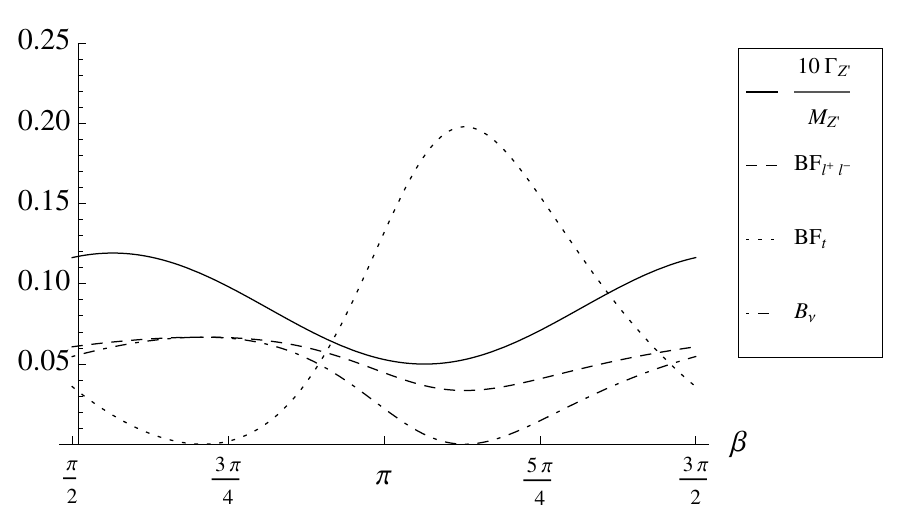}
\includegraphics[width=3in]{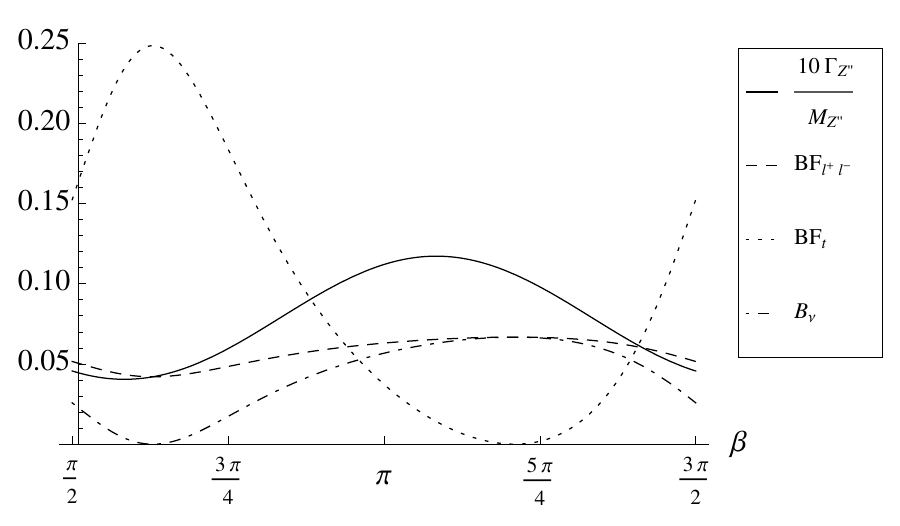}
\caption{\label{bf}Total width, normalized by the mass, and branching fractions to lepton, top, and neutrino pairs for the lower mass eigenstate $Z'$ (left) and the higher mass eigenstate $Z''$ (right). For both, we include only decays into SM fermions. }
\end{figure}

We now consider $E_6$ grand unification scenarios in more detail. An $E_6$ gauge group can be broken down into either a rank-5 or rank-6 subgroup. In the rank-5 case, this leads to one additional $Z'$ boson, the $Z'_{\eta}$ discussed above. In the rank-6 case, there are two additional $Z'$s, corresponding to the additional $U(1)_{\psi}$ and $U(1)_{\chi}$ groups in Eq.~(\ref{breaking chain}). The mass eigenstates are $Z'$ and $Z''$ of 
Eq.~(\ref{beta}).
We justifiably ignore small mixings of the $Z'$ bosons with the SM $Z$. Often, the $Z''$ is assumed to be very heavy, leading to an effective rank-5 group, as was the case in the models we considered in Section~\ref{Sim}.  In Fig.~\ref{bf}, we show the branching fractions and total width of each additional boson as a function of the mixing angle $\beta$. 

In the rank-6 case, the masses of the $Z'$ and $Z''$ are related by~\cite{Barger:1986nn}, 
\begin{equation}
\label{mass ratio}
\left(\frac{M_{Z'}}{M_{Z''}}\right)^2 = \left(\frac{\cos{\beta}+\sqrt{15} \sin{\beta}}{\sqrt{15} \cos{\beta}-\sin{\beta}} \right) \left(\frac{\cos{\beta}}{\sin{\beta}} \right)\,.
\end{equation}
This relation assumes that the $U(1)'$ symmetry breaking scale is much higher than the electroweak scale. Since the experimental lower bound on the $Z'$ mass is currently about 2~TeV for $E_6$ models, and there are tight limits on mixing with the SM $Z$, this is a justified assumption.

Requiring that the $Z''$ be heavier than the $Z'$, and that both masses are positive leads to the condition
\begin{equation}
\label{beta range}
-\sqrt{15}/4 \leq \cos{\beta} \leq 0
\end{equation}
Notice that the $Z'_{\eta}$, with $\cos{\beta} = \sqrt{5/8}$ is excluded from the range in Eq.~(\ref{beta range}). Additionally, both the $Z'_{\psi}$ and $Z'_{\chi}$, with $\cos{\beta} = 1$ and $\cos{\beta} = 0$, respectively, have $M_{Z'} \ll M_{Z''}$. Therefore, we would not expect the LHC to detect a heavier mass eigenstate for the three $E_6$ models considered in Section~\ref{Sim}.
 
\begin{figure}[!t]
\includegraphics[width=2.5in]{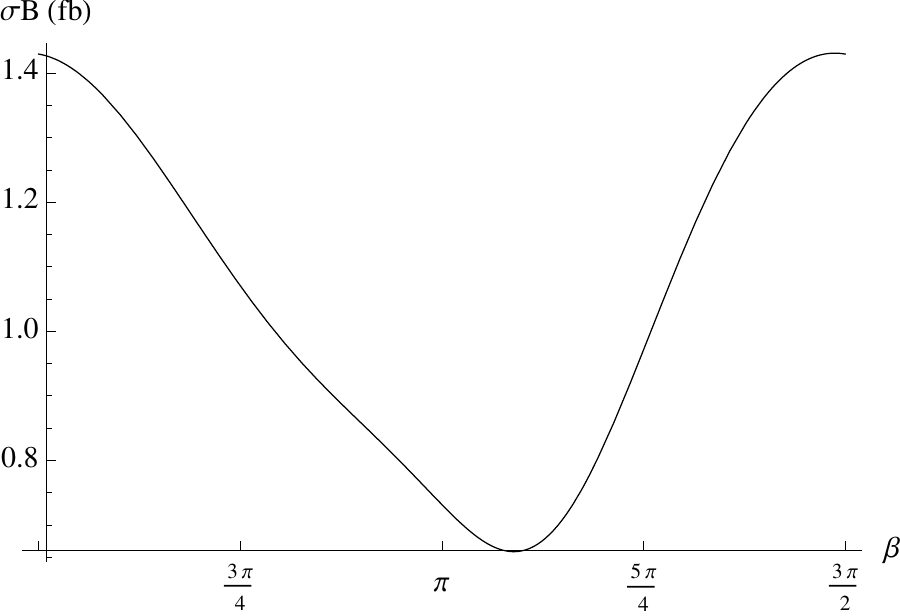} 
\caption{\label{xsec_beta_z2}Integrated peak cross section for the process $pp \rightarrow Z' \rightarrow l^+l^-$ as a function of mixing angle $\beta$, with $M_{Z'} = 2.5$~TeV at LHC14.} 
\end{figure}

\begin{figure}[!h]
\includegraphics[width=3.5in]{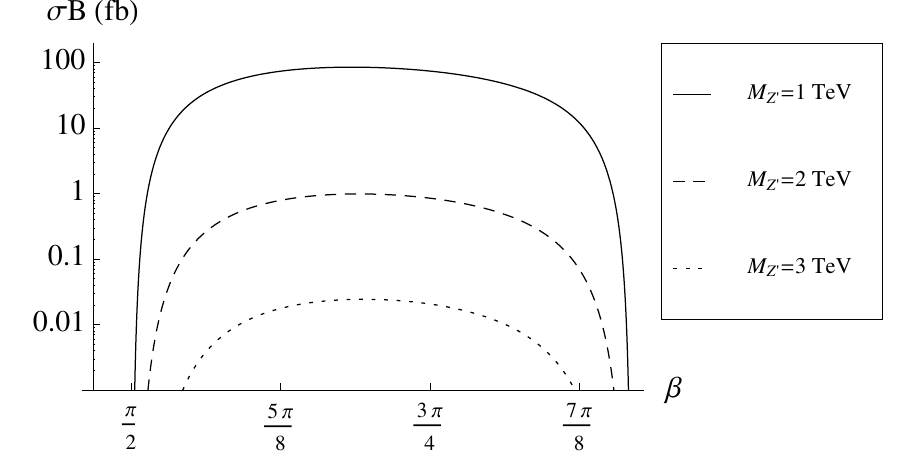}
\caption{\label{xsec_beta_z3}Integrated peak cross section for the process $pp \rightarrow Z'' \rightarrow l^+l^-$ at LHC14 versus mixing angle $\beta$.  The mass of the heavier eigenstate is determined by the $Z'$ mass and the mixing angle.}
\end{figure}

Using the empirical equations determined in Section~\ref{fits}, we can now calculate the cross section for the $Z'$ and $Z''$.  Both cross sections are a function of just two free parameters: $M_{Z'}$ and $\beta$.  In Fig.\  \ref{xsec_beta_z2}, we show the integrated peak cross section for $Z'$ production and decay at LHC14 versus mixing angle. In Fig.\  \ref{xsec_beta_z3}, we plot the same quantity for the $Z''$ over the allowed range of mixing angles. Here we see that if a $Z'$ were to be discovered in the mass range of $1-2.5$~TeV or so, the higher mass $Z''$ could be accessible at the LHC as well, within a certain range of mixing angles.

\section{\label{DMZ'} Dark Matter Interactions}

There are a number of scenarios in which a $Z'$ boson can serve as a dark matter mediator~\cite{An:2012va, Buckley:2011mm, Cline:2011uu,  Petraki:2011mv, Gondolo:2011eq}. These models often include mixing between the gauge bosons, leading to small effective couplings between dark matter and SM fermions~\cite{Heeck:2011md, Babu:1997st, Frandsen:2011cg, Chun:2010ve, Cassel:2009pu, Hook:2010tw, Mambrini:2010dq}. We study this possibility, paying particular attention to the possibility of isospin-violating dark matter scattering, which occurs naturally in the case of a $Z'$ mediator.

We now consider a model with a new $U(1)'$ and a new Dirac fermion that is charged only under $U(1)'$ -- this fermion will serve as our dark matter candidate. Interactions between the dark matter and SM particles are achieved through the kinetic and mass mixing of the new $Z'$ boson with the SM $Z$. The Lagrangian in this case is~\cite{Babu:1997st}
\begin{align*}
{\cal L} = &{\cal L}_{SM } - \frac{1}{4}\hat{Z'}_{\mu\nu}\hat{Z'}^{\mu\nu}+\frac{1}{2}M_{Z'}^2\hat{Z'}_{\mu}\hat{Z'}^{\mu} \\
&-\hat{g'}\sum_i \bar{\psi_i}\gamma^{\mu}(f^i_V-f^i_A\gamma^5) \bar{\psi_i}\hat{Z'_{\mu}} - \frac{\sin{\epsilon}}{2}\hat{Z'}_{\mu\nu}\hat{B}^{\mu\nu}+\delta M^2\hat{Z'}_{\mu}\hat{Z}^{\mu}\,.
\end{align*}
Here $\sin\epsilon$ and $\delta M^2$ parameterize the kinetic and mass mixing between the $Z'$ and the $Z$. As usual, $\hat{B}_{\mu\nu}$, $\hat{W}_{\mu\nu}$ and $\hat{Z'}_{\mu\nu}$ are the field strength tensors for $U(1)_Y$, $SU(2)_L$ and $U(1)'$. $\psi_i$ are the fermion fields (including the dark matter), and $f_V^i$ and $f_A^i$ are the vector and axial charges of the fermions under $U(1)'$. For simplicity, we consider the case where all SM fermions have  $f_V^i = f_A^i = 0$. This choice leads to rather weak couplings between SM fermions and the new $Z'$, which avoids current LHC bounds on $Z'$ production.  $f_V^{\chi}$ must be nonzero to allow for spin-independent scattering of dark matter on nuclei.

We define two additional parameters for convenience:
\begin{align}
\delta &= \frac{\delta M^2}{M_{\hat Z}^2}\,, \\ 
\tan 2\xi &=  \frac{-2\cos \epsilon  (\delta  + \hat{s}_W \sin \epsilon)}
{M_{\hat{Z}'}^2/M_{\hat{Z}}^2 - \cos^2 \epsilon+\hat{s}_W^2 \sin^2 \epsilon + 2 \delta  \sin \epsilon}\,,
\end{align}
where $\hat{s}_W$ is the sine of the weak mixing angle.
The  $Z'$ couplings to DM and the shifts in the $Z$ couplings are proportional to $\xi$. Since $\xi$ is approximately proportional to $M_{Z'}^{-2}$ for $\delta,\epsilon \ll \frac{M_{Z'}}{M_Z}$, these couplings must be small for heavy $Z'$s. 

The physical states $A_{\mu}$, $Z_{\mu}$, and $Z'_{\mu}$ are obtained through two sequential transformations. First, we diagonalize the field strength tensors; then, after $SU(2)\times U(1)$ breaking, we diagonalize the resulting mass matrices. After these transformations, the physical states are related to the original (hatted) states by
\begin{align}
A_{\mu} &= \hat{A}_{\mu} + \hat{c}_W \sin{\epsilon} \, \hat{Z'_{\mu} }\,,  \nonumber \\
Z_{\mu} &=  \cos{\xi} (\hat{Z}_{\mu}  - \hat{s}_W \sin{\epsilon} \hat{Z'_{\mu} })+\sin{\xi} \cos{\epsilon} \hat{Z'_{\mu} }\,, \\
Z'_{\mu} &= \cos{\epsilon} \cos{\xi} \hat{Z'_{\mu} } - \sin{\xi}(\hat{Z_{\mu} } -\hat{s}_W \sin{\epsilon} \hat{Z'_{\mu} } )\,. \nonumber
\end{align}

We can write the couplings of the physical states to fermions in terms of the oblique parameters, $S$, $T$, and $U$, and the ``physical" weak mixing 
angle~\cite{Babu:1997st}:
\begin{align}
\label{couplings}
g_{fZ}^V &= \frac{e}{2 s_W c_W}\left(1 + \frac{\alpha T}{2}\right)(T_3^i - 2 Q^i s_*^2)\,, \nonumber \\
g_{fZ}^A &=\frac{e}{2 s_W c_W} \left(1 + \frac{\alpha T}{2}\right)T^i_3\,, \nonumber \\
g_{fZ'}^V &= \frac{e}{2 s_W c_W}\left(1 + \frac{\alpha T}{2}\right)\left(\tilde{s}(T_3^i - 2 Q^i) \tan{\epsilon}  - (T_3^i - 2 Q^i s_*^2) \xi\right)\,, \nonumber \\
g_{fZ'}^A &= \frac{e}{2 s_W c_W}\left(1 + \frac{\alpha T}{2}\right) \left( \tilde{s} T^i_3 \tan{\epsilon}  -T^i_3 \xi\right)\,,  \nonumber\\
g_{\chi Z}^V &= \xi f^\chi_V\,, \nonumber\\
g_{\chi Z'}^V &= f^\chi_V\,,
\end{align}
where \begin{equation}
s_*^2 = s_W^2 + \frac{1}{c_W^2 - s_W^2} (\frac{1}{4} \alpha S - c_W^2 s_W^2 \alpha T)\,,
\end{equation}
and 
\begin{equation}
\tilde{s} = s_W + \frac{s_W^3}{c_W^2 - s_W^2}(\frac{1}{4 c_W^2} \alpha S -\frac{1}{2} \alpha T)\,.
\end{equation}
The contributions to $S$ and $T$ due to the $Z'$ are, to second order in $\xi$~\cite{Babu:1997st}, 
\begin{equation}
 \alpha S = 4 \xi c_W^2 s_W \tan{\epsilon} \,,
\end{equation}
\begin{equation}
 \alpha T = \xi^2 (\frac{M_{Z'}^2}{M_Z^2} - 1) + 2 \xi s_W \tan{\epsilon}\,. 
\end{equation}
\begin{figure}
\includegraphics[width=3in]{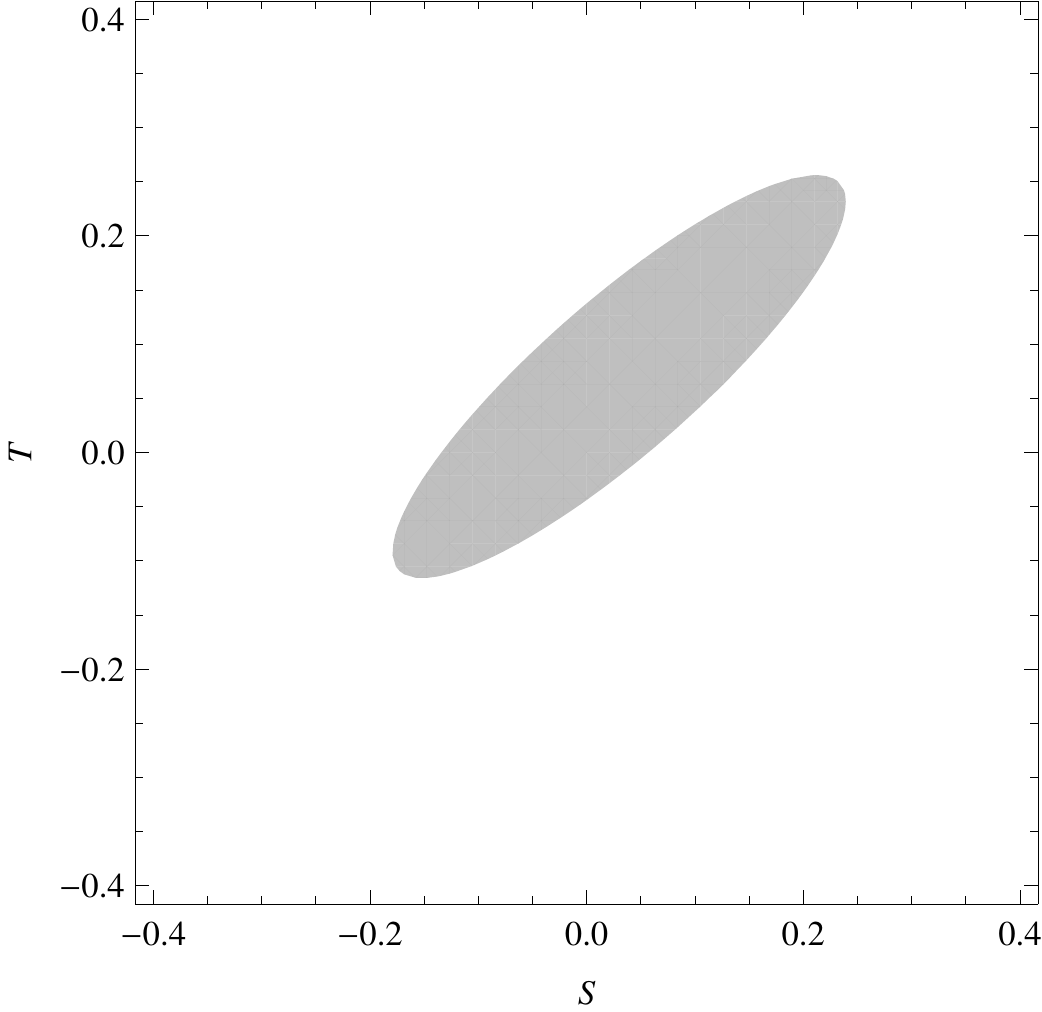}
\includegraphics[width=3in]{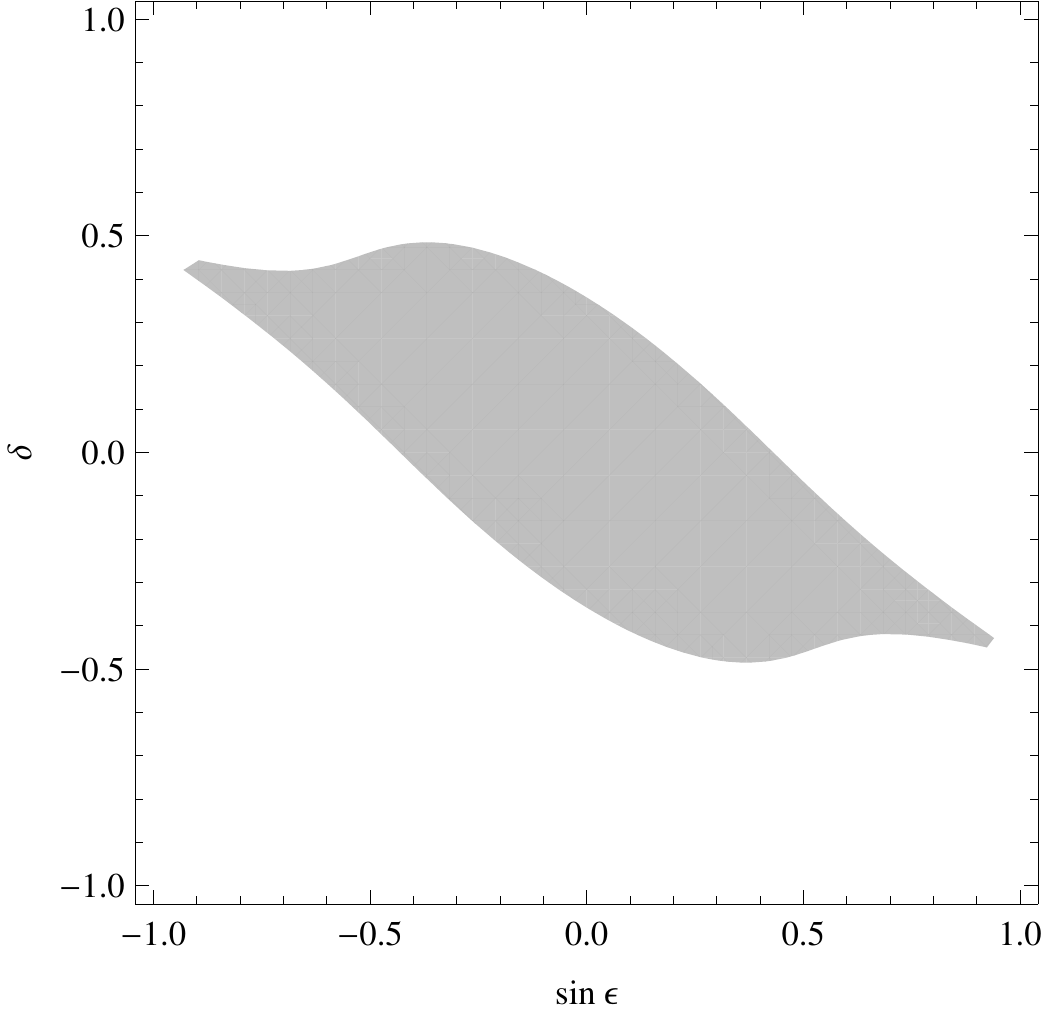}
\caption{\label{st_pdg}Left: The 90\% C.L. allowed region for $S$ and $T$ based on the global data for $M_H = 117$~GeV~\cite{Nakamura:2010zzi}. Right: This region translated into a 90\% CL region for $\epsilon$ and $\delta$ for $M_{Z'} = 1000$~GeV.}
\end{figure}
$S$ and $T$ are constrained by fits to the global electroweak data, as shown in Fig.~\ref{st_pdg}~\cite{Nakamura:2010zzi}. The best fit values for an assumed Higgs boson mass of $M_H = 117$~GeV are 
\begin{align*}
S = 0.03 \pm 0.09\,, \\
T = 0.07 \pm 0.08\,,
\end{align*}
with a strong correlation of $87\%$. The $S$ and $T$ values would change very little for a Higgs mass of 125~GeV, as may be suggested by recent LHC observations~\cite{Chatrchyan:2012tx, Collaboration:2012si}. The constraints on $S$ and $T$ can be translated into constraints on the $Z'$ mixing angles (and consequently the couplings and cross sections) with a simple Monte Carlo. The result is shown in Fig.~\ref{st_pdg}. Note that the kinetic mixing angle $\epsilon$ can be quite large, assuming that $\delta$ is small enough.  

By sampling values in the region of allowed $\epsilon$ and $\delta$, we can determine the range of possible values for the dark matter scattering cross section and the LHC $Z'$ production cross section. These can then be compared to the results of experimental searches to further restrict the mixing angles, $Z'$ mass, and $f_n/f_p$ (defined below).

For direct detection experiments, we are interested in interactions between dark matter and atomic nuclei~\cite{Frandsen:2011cg}. To determine this cross section, we start with an effective dark matter-quark coupling given by
\begin{equation}
b_f^{V,A} = \frac{g_{\chi Z'}^{V,A} g_{f Z'}^{V,A}}{M_{Z'}^2}+\frac{g_{\chi Z} ^{V,A}g_{f Z}^{V,A}}{M_{Z}^2}\,,
\end{equation}
 which leads to DM-nucleon couplings of
\begin{align}
f_n &= 2 b_d^V + b_u^V\,, \\ \nonumber
f_p &= b_d^V +2 b_u^V\,.
\end{align}

\begin{figure}
\includegraphics[width=3in]{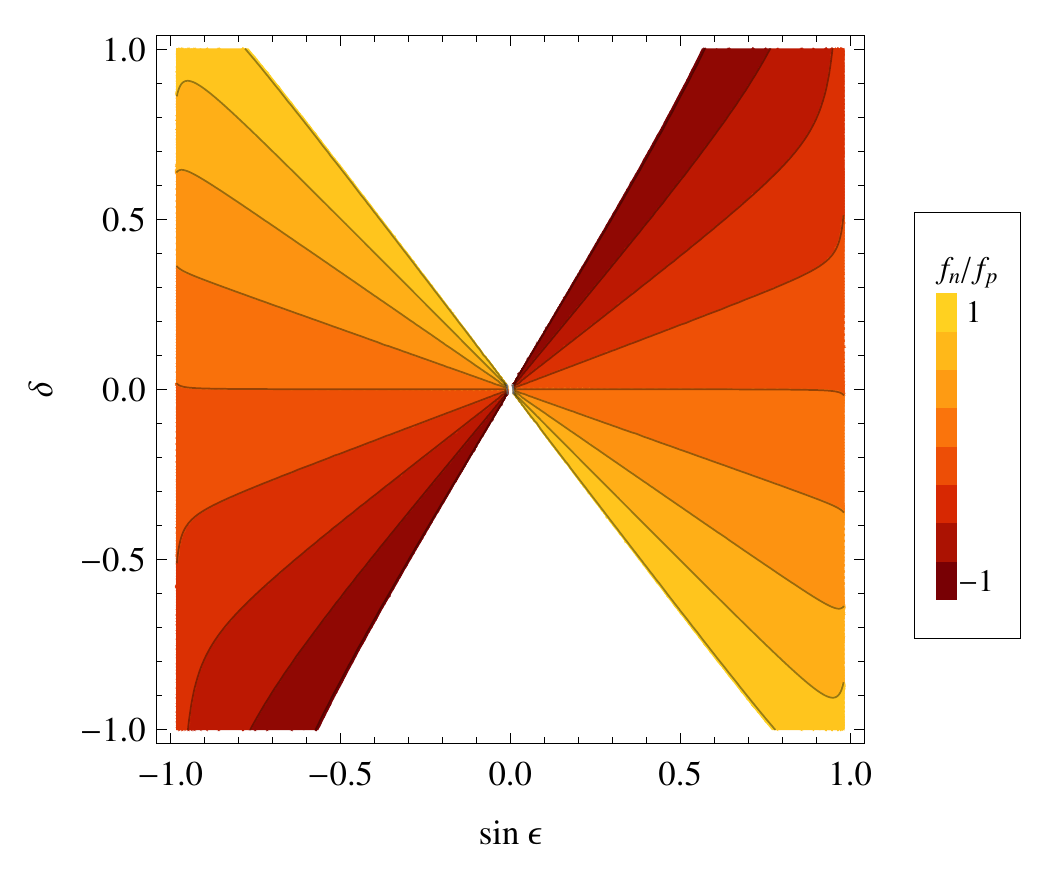}
\includegraphics[width=3in]{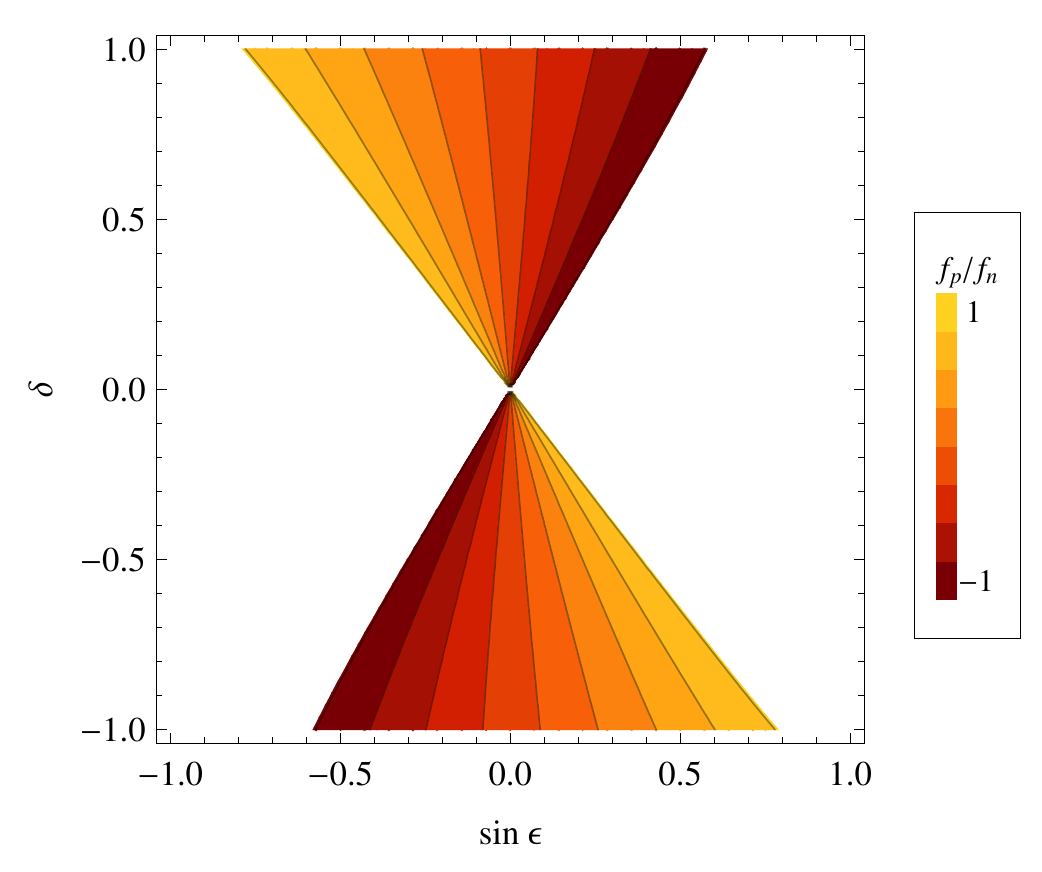}
\caption{\label{fn_fp} Contour plots of $f_n/f_p$ and  $f_p/f_n$ as a function of $\epsilon$ and $\delta$. No limits are placed on the values of $\epsilon$ and $\delta$.}
\end{figure}

Using these expressions, we can determine $f_n/f_p$ as a function of $\epsilon$, $\delta$, and $M_{Z'}$. Contour plots of $f_n/f_p$ and $f_p/f_n$ as functions of $\epsilon$ and $\delta$ are shown in Fig.~\ref{fn_fp}. Comparing with Fig.~\ref{st_pdg}, we see that $S$  and $T$ place no limit on the value of $f_n/f_p$, though the mixing angles are more tightly constrained for some values of  $f_n/f_p$ than others. Therefore, the limit on the dark matter scattering cross section will vary significantly as a function of  $f_n/f_p$.

Finally, we can write the dark matter-nucleus spin-independent scattering cross section:
\begin{equation}
\label{sigmaN}
\sigma_A = \frac{\mu_{A}^2}{\mu_{p}^2}\left( Z + (A -Z) \frac{f_n}{f_p} \right)^2 \sigma_p\,,
\end{equation}
where $\sigma_p$ is the spin-independent DM-proton cross section, 
\begin{equation}
\sigma_p = {{\mu_{p}^2 f_p^2}\over{64\pi}}\,,
\end{equation}  
and $Z$ and $A$ are the atomic and mass numbers of the detector material, $\mu_{A}$ is the reduced mass of the dark 
matter-nucleus system, and $\mu_{p}$ is the reduced mass of the dark matter-proton system~\cite{Feng:2011vu}.

\begin{figure}
\includegraphics[width=3in]{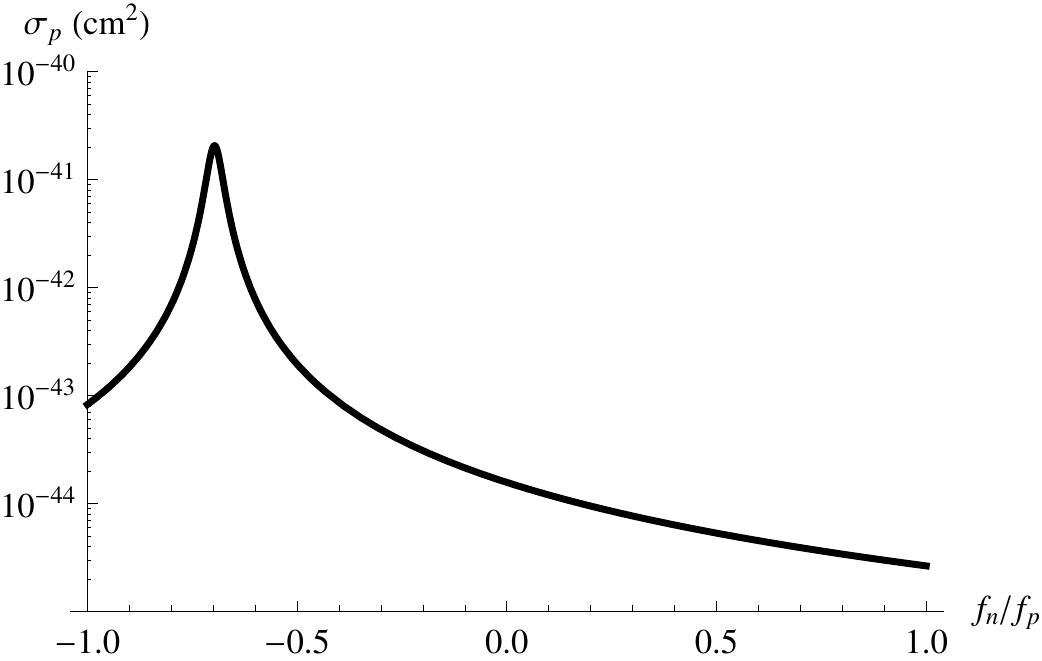}
\caption{\label{xenonlimit} The 90\% C.L. upper bound on the spin independent dark matter-proton scattering cross section from XENON100 as a function of 
$f_n/f_p$.} 
\end{figure}

It is common to present the spin-independent cross section for a nucleus with Z protons in terms of the cross section for scattering off a single nucleon, making the assumption that $f_n=f_p$. To compare a model with data, we must also account for possible isospin violation. To convert the nuclear cross section to a proton cross section $\sigma_p$, we multiply by~\cite{Feng:2011vu}
\begin{equation}
\label{ratio}
F_Z = \frac{\sum_i \eta_i \mu_{A_i}^2 A_i{}^2}{\sum_i \eta_i \mu_{A_i}^2 [Z + (A_i-Z) f_n/f_p]^2}\,,
\end{equation}
This factor is derived from Eq.~\eqref{sigmaN} by summing over all stable isotopes of atomic number $Z$, weighted by their natural abundances $\eta_i$.  In Fig.~\ref{xenonlimit}, we show the effect of isospin violation on the xenon (Z = 54) cross section, normalized to the current limit from XENON100, $\sigma_p < 2.7 \times 10^{-45}$~cm$^2$ at $f_n/f_p = 1$ and $M_\chi = 100$~GeV~\cite{Aprile:2012nq}. Isospin violation can relax the bound by several orders of magnitude, with the least stringent bound occurring around $f_n/f_p = -0.7$~\cite{Feng:2011vu}.
%

\begin{figure}
\includegraphics[width=3in]{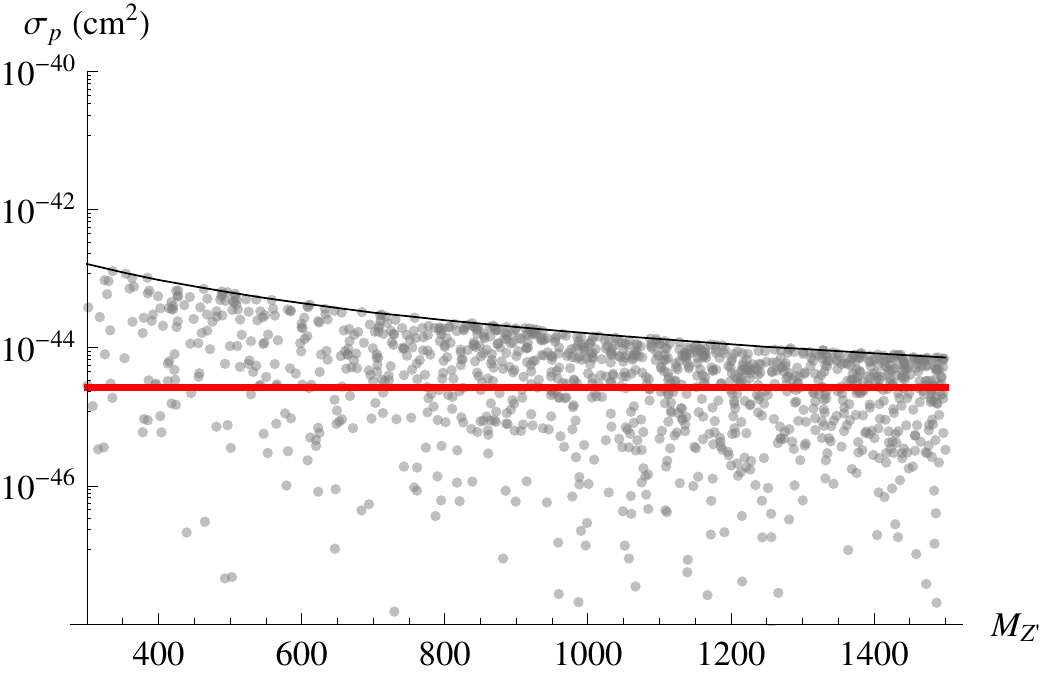}
\caption{\label{sigmap_mz} The cross section fro DM scattering on protons, $\sigma_p$ versus $Z'$ mass for $f_n/f_p=1$.  The thick red line at
$\sigma_p=2.7 \times 10^{-45}$~cm$^2$ is the XENON100 limit~\cite{Aprile:2012nq}; dots are the cross sections corresponding to pairs  ($\epsilon$, $\delta$) sampled uniformly over the allowed values. The dark matter mass and coupling to the $Z'$ are set to $M_\chi = 100$~GeV and $g_\chi = 1$, respectively.} 
\end{figure}

\begin{figure}
\includegraphics[width=3in]{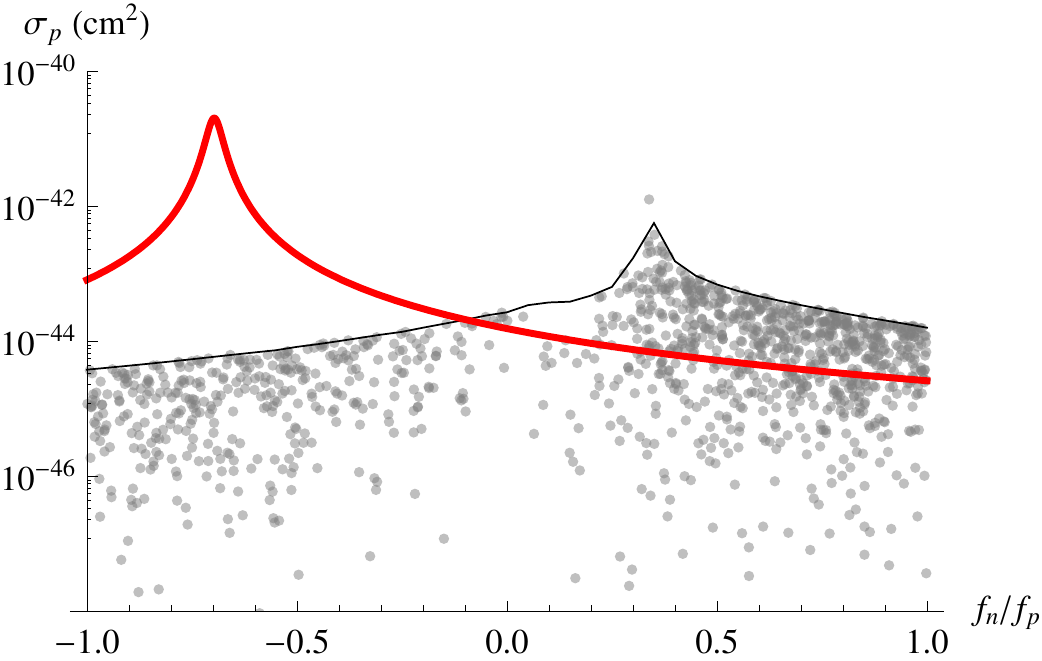}
\caption{\label{sigmap_fnfp} $\sigma_p$ versus $f_n/f_p$ mass for $M_Z' = 1000$~GeV.  The thick red line is the XENON100 upper bound; dots are the cross sections corresponding to pairs  ($\epsilon$, $\delta$) sampled uniformly through the allowed region shown in Fig \ref{st_pdg}. The dark matter mass and coupling to the $Z'$ are set to $M_\chi = 100$~GeV and $g_{\chi} = 1$, respectively.} 
\end{figure}

With the XENON bound generalized to all values of $f_n/f_p$, we can use it to place limits on the model parameters. We start by choosing a random sample of points $(\epsilon,\xi)$ within the allowed region shown in Fig.~\ref{st_pdg} and then calculate the proton SI cross section for each point. In 
Figs.~\ref{sigmap_mz} and~\ref{sigmap_fnfp}, we show the dependence of the cross section on $f_n/f_p$ and $M_{Z'}$, respectively. 
From Fig.~\ref{sigmap_fnfp}, we see that the largest cross sections occur for $f_n/f_p \approx 0.35$, while the lowest occur for $f_n/f_p < -0.5$, which is where the XENON bound is the most relaxed. For both figures, we have set $g_{\chi}=1$ and $M_{\chi}=100$~GeV, which in general are free parameters. However, they have no impact on the qualitative features of the distributions, as they only enter the overall normalization factor. They  influence our ability to place limits on the other parameters; in particular, the $g_{\chi}^2$ dependence of the cross section means that all constraints 
can be evaded by choosing a small enough coupling. For $M_{Z'} = 1000$~GeV and $M_\chi = 100$~GeV, the XENON bounds are evaded for all values of $f_n/f_p$ with $g_{\chi} =0.58$.

\begin{figure}[!t]
\includegraphics[width=3in]{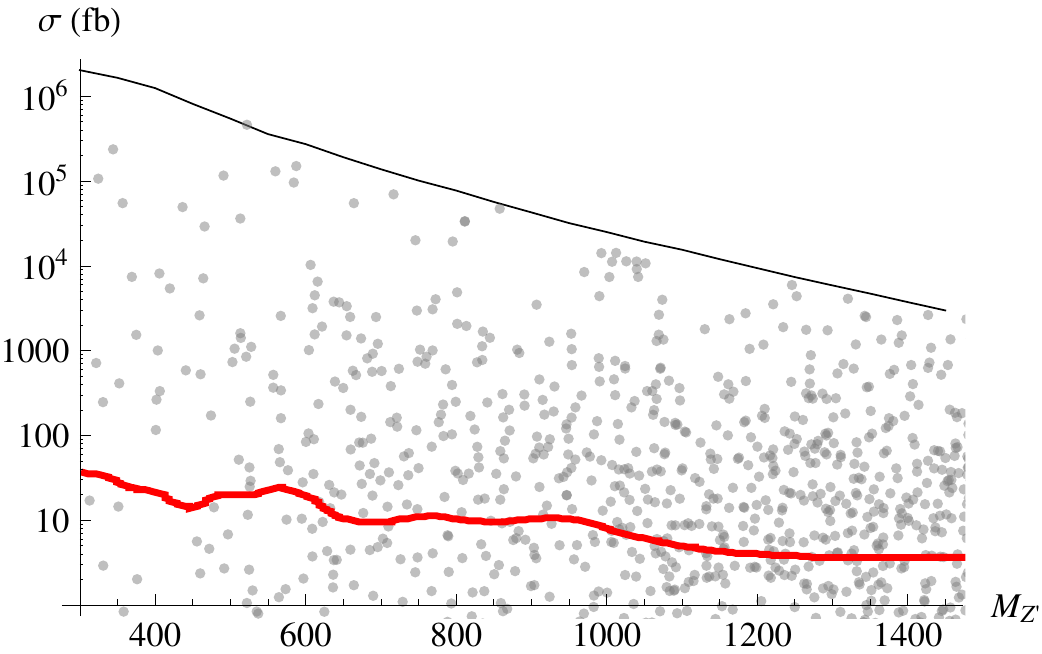}
\caption{\label{LHC} $\sigma(pp \rightarrow Z' \rightarrow l^+l^-)$ versus $M_Z'$. The thick red line is the most recent ATLAS limit~\cite{Collaboration:2011dca}; dots are the cross sections corresponding to pairs  ($\epsilon$, $\delta$) sampled uniformly through the allowed region shown in Fig \ref{st_pdg}.} 
\end{figure}

Collider searches can also help constrain the mixing angles for a dark $Z'$, as the $Z'$ acquires small couplings to SM fermions via mixing effects \cite{Frandsen:2012rk}. $Z'$ production and decay to leptons at the LHC depends primarily on the mixing angles and the $Z'$ mass, with only a small dependence on the dark matter properties through the decay width of the $Z'$. Using the couplings in Eq.~(\ref{couplings}), we can apply the analysis of Section~\ref{fits} to calculate the cross section for $Z'$ production at the LHC. We find the ATLAS predictions of the cross section for the various $Z'$ models given in Fig.~2 of
Ref.~\cite{Collaboration:2011dca} are well parameterized by 
\begin{equation}
\sigma B= (2200~{\rm fb}) \ r_{Z'}^{12} \left( \frac{1}{r_{Z'}}-1\right)^{15} \left[p \, c_u + (1-p) \, c_d\right]\,.
\end{equation}
With this equation, we can use the current ATLAS limits to restrict the parameter space. Since there are no direct couplings between the $Z'$ and the SM, the lower bound on the $Z'$ mass is much less stringent than for the models considered earlier. In Fig.~\ref{LHC}, we show the upper bound on $\sigma(pp \rightarrow Z' \rightarrow l^+l^-)$ set by the $S$ and $T$ parameters for $\sqrt{s} = 7$~TeV.

\begin{figure}[!t]
\includegraphics[width=3in]{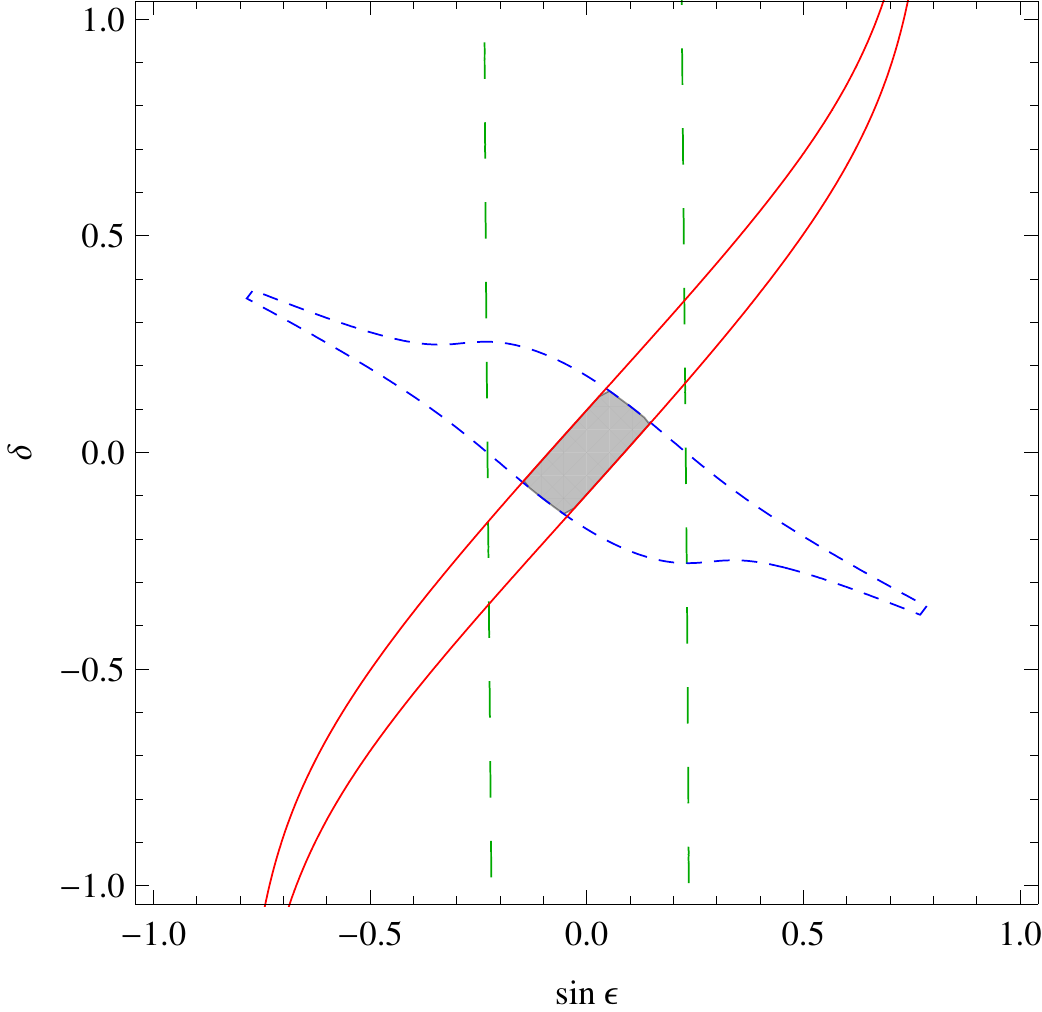}
\includegraphics[width=3in]{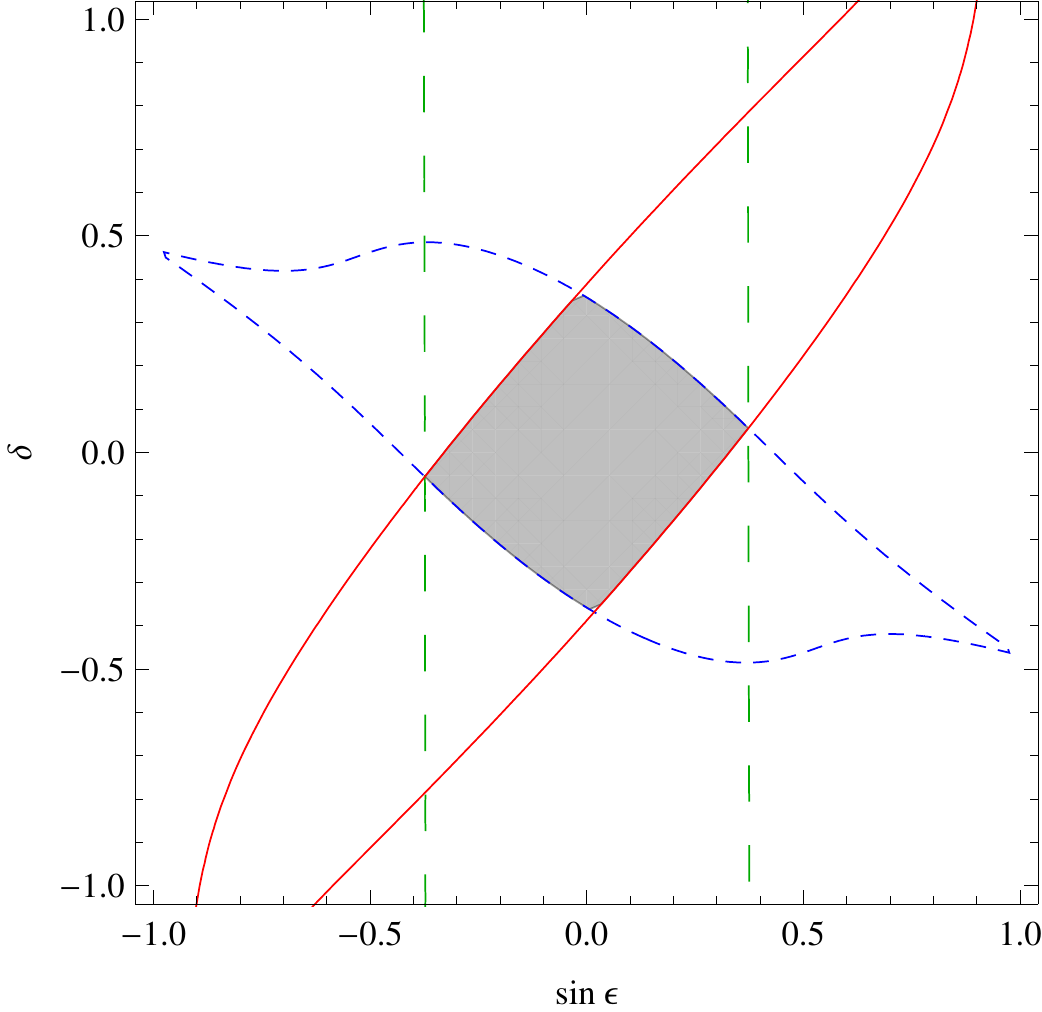}
\caption{\label{limits} Derived limits on $\epsilon$ and $\delta$ for $M_{Z'}= 500$~GeV (left) and $M_{Z'}= 1000$~GeV (right). The dotted blue lines represent the limit from global electroweak data (90$\%$ C.L.), the solid red lines are the limit from XENON100 (90$\%$ C.L.), and the dashed green lines are the limits from ATLAS at $\sqrt{s} =7$~TeV for $\mathcal{L} \approx 1$~fb$^{-1}$ (95$\%$ C.L.).}
\end{figure}

Finally, we can combine the limits from XENON100 and ATLAS to constrain the model parameter space in terms of $\epsilon$, $\delta$, and $M_{Z'}$. The results are shown in Fig.~\ref{limits}, again with $M_\chi = 100$~GeV and $g_\chi = 1$.  The electroweak, dark matter, and LHC data provide complimentary bounds, with $S$ and $T$ more strongly limiting the degree of mass mixing, while XENON and ATLAS provide more stringent bounds on kinetic mixing. The bounds on $\delta$ and $\epsilon$ relax as $M_{Z'}$ increases and the $Z$ and $Z'$ decouple. Note that the experiments report their results at different confidence levels, so these regions are not confidence regions; they are simply indicative of the parameter space available.

%

The invisible decay width of the $Z$ and the muon anomalous magnetic moment have also been used to constrain the mixing parameters $\epsilon$ and $\delta$ ~\cite{Chun:2010ve}, but they are less restrictive than $S$ and $T$ for $M_{Z'}$ in the range $300-1500$~GeV. 

If $M_{Z} > M_{\chi}$, the invisible decay width of the $Z$ must be considered. The decay width is proportional to $\xi^2$, so the experimental 1.5 MeV bound~\cite{Nakamura:2010zzi} is avoided as long as $\xi < 0.95$. 

There are also corrections to the muon anomalous magnetic moment~\cite{Chun:2010ve},
\begin{equation}
\delta a_{\mu} \approx \frac{\alpha \xi^2}{3 \pi c_W^2 s_W^2}\frac{m_\mu^2}{M_{Z'}^2}\,.
\end{equation}
\nopagebreak
However, the contributions from the $Z'$ are very small, because of the dependence on $M_{Z'}^{-2}$ and $\xi^2$. The current experimental limit is $\delta a_{\mu} \approx 3\times 10^{-9}$~\cite{Nakamura:2010zzi}, which is several orders of magnitude larger than the expected contribution except for very small $M_{Z'}$. 

\section{Summary}
Our customization of the FEWZ simulation code allows for extensive studies of $Z'$ production and decay at the LHC at NLO and NNLO. Using the results of our simulation for representative benchmark models, we derived an empirical formula for the double differential cross section $\frac{d^2\sigma}{dy\, d\cos{\theta}}$. This formula can be used to study broad classes of models easily and  to determine the $Z'$ couplings to fermions without prior knowledge of the underlying model.  In a model-dependent analysis, it can also be used to set limits on model parameters. In the case of the $E_6$-derived models, without accounting for systematic uncertainties, we find that the mixing angle $\beta$ may be determined within 0.1$\pi$ with 1~ab$^{-1}$ of data for a mass of 2.5~TeV.  For 
$E_6$ models we also showed that within a range of mixing angles, two extra neutral gauge bosons should be within the reach of the LHC. 

Finally, we considered a more general $Z'$ model with kinetic and mass mixing, which has interesting implications for dark matter detection and hidden sector theories. Even if the SM is uncharged under a new hidden sector $U(1)'$, mixing could induce couplings strong enough that the $Z'$ could be produced at the LHC and mediate dark matter scattering on nuclei, without violating limits from global electroweak data. In the case of a heavy 
$Z'$ ($M_{Z'} \gg M_{Z}$) we find that the limits on the $S$ and $T$ parameters can be combined with XENON100 and LHC data to restrict the range of allowed mixing angles. For a relatively light $Z'$ at 500~GeV, the mass and kinetic mixing parameters $\delta$ and $\epsilon$ must both be less than about 0.2. For heavy $Z'$s, these mixing parameters are unrestricted.

\section*{Acknowledgments}
DM thanks the Galileo Galilei Institute for Theoretical Physics for its hospitality during the completion of this work.  This research was supported by the DoE under Grant Nos. DE-FG02-95ER40896 and DE-FG02-04ER41308, by the NSF under
Grant No. PHY-0544278, and by the Wisconsin Alumni Research Foundation.

\appendix*
\section{\label{appendix}FEWZ Customization}
The FEWZ code allows for an extensive analysis of the SM Drell-Yan process, including NNLO effects and the influence of phase space cuts.  With a few modifications, its features can be used to study $Z'$ production as well. The details of the original simulation are provided in Refs.~\cite{Gavin:2012kw, Gavin:2010az}.

The QCD factorization theorem allows us to write the $Z'$ production cross section in terms of the partonic cross section and the proton structure functions as follows:
\begin{equation}
\label{diffxsect}
d\sigma = \sum_{ij} \int dx_1 dx_2 f_i^{h_1}(x_1) f_j^{h_2}(x_2) d\sigma_{ij \rightarrow l_1 l_2}(x_1 , x_2)\,. \nonumber
\end{equation}
Since the addition of a $Z'$ affects only the partonic cross section, and not the structure functions, we do not need to alter the Monte Carlo portion of FEWZ.

The parameters of the $Z'$ model are specified in the input file for each run.  For each model, the user sets the mass, total width, partial width to leptons, and couplings of the $Z'$. The input file also includes the SM parameters and kinematic cuts.  We also include a switch to turn on and off the $Z'$ contribution, so that calculations of the SM background can still be done. 

The $Z'$ parameters are read into FEWZ and used to calculate ``weights" (related to the partonic cross section of various subprocesses) for the integration routine. Preserving the structure of the SM calculation, we insert additional $Z'$ contributions to the partonic cross section. This includes the interference terms mentioned in Eq.~(\ref{xsec_interference}):
\begin{equation}
\sigma(Z^\prime,X)
=\frac{g_{Z'} g_X}{2\pi}\,
\left( \frac{ z_{l_L}\,z^{X}_{l_L} + z_{l_R}\,z^{X}_{l_R} }{288}\right)
\,\frac{(Q^2-M_{Z^\prime}^2)\,(Q^2-M_{X}^2)+M_{Z^\prime}\,M_X\,
\Gamma_{Z^\prime}\,\Gamma_{X}}
{
\left[ \left( Q^2-M_{Z^{\prime}}^2 \right)^2
+ M_{Z^{\prime}}^2\,\Gamma_{Z^{\prime}}^2\right]\,
\left[ \left( Q^2-M_{X}^2 \right)^2 + M_{X}\,\Gamma_{X}^2 \right]
}\,, \nonumber
\end{equation}
where $X = \gamma,\,Z$. Hereafter, the Monte Carlo integration proceeds without alteration. 

\bibliographystyle{utphys.bst}
\bibliography{zprimebib}

\end{document}